\documentclass[man, a4paper]{apa6}
\usepackage{times} 		    

\usepackage{amsmath} 	 
\usepackage{geometry}    
\usepackage{longtable}   
\usepackage{pdflscape}   
\usepackage{rotating}    
\usepackage{setspace} 	 
\usepackage{etoolbox}    
\patchcmd{\maketitle}{\vspace*{1in}}{}{}{}
\usepackage[flushleft]{threeparttable}  

\usepackage[utf8]{inputenc}
\usepackage{csquotes,xpatch}
\usepackage[american]{babel}
\usepackage[style = apa, sortcites = true, sorting = nyt, backend = biber]{biblatex}
\DeclareLanguageMapping{american}{american-apa}
\usepackage{hyperref}
\hypersetup{
	pdftitle   = {Assessing Individual Change Processes},
	pdfauthor  = {Jean-Paul Fox},
	pdfsubject = {Bayesian Covariance Structure Modelling (BCSM),  negative clustering effects, personalized intervention, multilevel modeling},
	colorlinks = true,
	citecolor  = black,
	urlcolor   = black,
	linkcolor  = black,
	filecolor  = black
}

\twoauthors{Jean-Paul Fox*}{Wouter A. C. Smink}
\leftheader{Fox \& Smink (2020)}

\twoaffiliations
{Research Methodology, Measurement \& Data Analysis \\ University of Twente}
{Research Methodology, Measurement \& Data Analysis \\ University of Twente}

\authornote{\noindent Please address all correspondence concerning this article to Jean-Paul Fox, Faculty of Behavior and Management Sciences, Department of Research Methodology, Measurement \& Data Analysis, University of Twente, P.O. Box 217, 7500 A.E. Enschede, The Netherlands. E-mail: \href{mailto:J.P.Fox@utwente.nl}{J.P.Fox@utwente.nl}.}

\title{Assessing an Alternative for `Negative Variance Components': A Gentle Introduction to Bayesian Covariance Structure Modelling for Negative Associations Among Patients with Personalized Treatments}

\shorttitle{Introducing BCSM For Negative Clustering Effects}

\abstract{The multilevel model (MLM) is the popular approach to describe dependences of hierarchically clustered observations. A main feature is the capability to estimate (cluster-specific) random effect parameters, while their distribution describes the variation across clusters. However, the MLM can only model positive associations among clustered observations, and it is not suitable for small sample sizes. The limitation of the MLM becomes apparent when estimation methods produce negative estimates for random effect variances, which can be seen as an indication that observations are negatively correlated.
A gentle introduction to Bayesian Covariance Structure Modelling (BCSM) is given, which makes it possible to model also negatively correlated observations. The BCSM does not model dependences through random (cluster-specific) effects, but through a covariance matrix. We show that this makes the BCSM particularly useful for small data samples.
We draw specific attention to detect effects of a personalized intervention. The effect of a personalized treatment can differ across individuals, and this can lead to negative associations among measurements of individuals who are treated by the same therapist. It is shown that the BCSM enables the modeling of negative associations among clustered measurements and aids in the interpretation of negative clustering effects. Through a simulation study and by analysis of a real data example, we discuss the suitability of the BCSM for small data sets and for exploring effects of individualized treatments, specifically when (standard) MLM software produces negative or zero variance estimates.
}


\begin{document}
	
\maketitle 		

\section{Introduction}
Data are so often plagued by observations that are correlated (i.e. clustered, not independently sampled) that it is difficult to overstate the importance of multilevel models. This family of statistical models aids researchers in understanding the clustered --or hierarchical-- structures in their data (i.e. `groups within groups', `non-independent data', or `hierarchical data'). In the multilevel modelling framework, dependences among observations are expressed as a covariance, which are modelled through a random effect, also known as a latent variable. The variance of the random effect determines the strength of the correlation among clustered observations. For a small variance, clusters are similar to each other and observations within a cluster do not correlate highly. With high random effect variance, the cluster-specific parameters show large differences and observations within each cluster are much more alike than those from different clusters.

However, modelling the clustering effect (i.e. magnitude of the --positive-- correlation between observations) as the variance of a random effect also introduces a great --and relatively unknown-- shortcoming of multilevel models. Multilevel models impose the restriction that within-cluster correlations should be positive, since the variance of a random effect is restricted to be positive. However, correlations are not restricted to positive values only, they can also be negative or zero. Indeed, although not widely known, negative correlations among clustered observations can also occur \parencite{Kenny2002}, but the multilevel model cannot assess these effects. {\color{black} Furthermore, independent of the sign and level of association, a cluster is a higher-level unit represented by lower-level observations, and care should be taken in analysing disaggregated (atomistic fallacy) or aggregated data (ecological fallacy).
}

The multilevel model describes within-cluster \textit{similarity} through a \textit{positive} random effect variance. From this perspective to describe within-cluster \textit{dissimilarity} with a multilevel model, a negative random effect variance would be required. This has led to interest in estimating and interpreting negative variance components and identifying negative clustering effects \parencite{Kenny2002, Molenberghs2007, Molenberghs2011, Pryseley2011, Oliveira2017, Verbeke2003, ElLeithy2016, Nelder1954}. These effects remain unknown to the sheer majority of scientific community. Furthermore, methods to apply multilevel modeling techniques for analysing within-cluster dissimilarities are limited, and have not been expanded to address more complex clustered data. \textcite{Molenberghs2011} discussed a marginal model representation of the random intercept model (by integrating out the random effect), and \textcite{Snijders1999} adjusted a multilevel model with correlated dummy variables to describe negative within-cluster dependence.

Recently, \textcite{Fox2017, Klotzke2019a}, and \textcite{Klotzke2019b}, developed a rigorous new Bayesian modeling approach for clustered data, referred to as \textit{Bayesian Covariance Structure Modelling} (BCSM). Based on multilevel modeling principles, BCSM can describe similarities and dissimilarities. We intend to give a gentle introduction to the BCSM here, and stress possibilities of the framework to deal with negatively and positively correlated observations. BCSM is a relatively simple and flexible covariance structure modelling approach, which avoids several restrictions of the popular multilevel models.

In short, in the BCSM approach, a dependence structure is not \textit{indirectly} modelled through random effect parameters. The dependence structure is \textit{directly} modelled by specifying a structured covariance matrix. This structured covariance matrix represents the correlations among clustered observations to account for the fact that the observations are not independently distributed. Both modelling approaches are discussed by considering the one-way random effects model (i.e. random intercept model). This relatively simple model is used as a vehicle to introduce the BCSM and its potential for modelling clustered data. Subsequently, BCSMs for more complex dependence structures, for instance a two-way (nested) structure, are described. The potential of the BCSM is supported by a straightforward Gibbs sampling method to estimate all model parameters, where (co)variance parameters can be directly sampled from inverse-gamma distributions.

We organized the remainder of the this paper as follows. We give a gentle introduction to modelling clustered data using random effects. Then, we introduce BCSM, emphasizing --not technical rigour, but-- understanding of the framework. We aim to convince those who are potentially interested in BCSM about the advantages of the approach by reporting on the results of our extensive simulation study, which shows that BCSM can --indeed-- detect positive as well as negative within-cluster dependence. In addition to that, we show that (very) small variance components of random effects (i.e. that are very close to zero) can be accurately estimated. We also demonstrate that BCSM can describe efficiently complex dependence structures with a few (co)variance parameters making it particularly useful for small data samples. We assess a real-data example, where differences in pre- and post-intervention depression scores between two treatment arms are examined, while accounting for a clustering by counsellors. The example illustrates why a negative clustering effect cannot be ignored, as these effects also occur in practice. Our overall goal is to not only discuss the statistical importance of negative clustering effects, but to show how negative effects should be interpreted. We will argue why clinical practitioners and psychotherapy researchers \parencite[and all others who are interested in knowing \textit{what works when for whom};][]{Norcross2011, Tasca2015, Smink2019_PLOS_One} are -- in fact-- interested in interpreting negative clustering effects. Finally, the specific features of BCSM are discussed, including its strengths and limitations.

\section{Modelling Clustered Data}

It is clear that various complex forms of clustering and hierarchical organisations arise naturally in a multitude of settings in psychological research. What all these settings have in common, is that --in their fundamental form-- each multilevel model consists out of a \textit{within} and a \textit{between} cluster component. With two levels,  the multilevel model (MLM) defines separate probability distributions for the clusters and for individuals within these clusters. Under the cluster-sampling design, clusters are assumed to be independently sampled from a population, and individuals are assumed to be independently sampled from each cluster. This two-stage sampling design is represented in the MLM, which specifies a probability distribution for the cluster-specific parameters (i.e. random effects) and a probability distribution for the lower-level observations. Following the properties of the two-stage sampling design, observations from individuals are assumed to be conditionally independently distributed given the cluster-specific (random effect) parameters.

Cluster-specific parameters (i.e. random effects) are often used to model clustered data and they are included in the mean regression component. When conditioning on the cluster-specific parameters, the observations within the cluster can be assumed to be independently distributed. In doing so, the assumption of independence is no longer violated, as the correlation of the clustered data is bypassed through inclusion of these cluster-specific parameters. This technique is used in the very popular models as hierarchical linear regression models \parencite{RaudenbushBryk2002}, random effect models \parencite{Longford1995}, multilevel models \parencites{Goldstein2011, SnijdersBosker2012}, and linear mixed effect models \parencites{McCulloch2008, Verbeke2009}. We use the term MLM to represent these type of (conditional) models. It is (also) good to note that this class of models can be extended further: in the latent class models \parencite{Vermunt2008}, or mixture models \parencite{McLachlan2000}, observations within each latent cluster are also assumed to be conditionally independently distributed given the cluster-specific parameters.

We use the one-way random effects model (i.e. random intercept model) to demonstrate the modelling of the within-cluster correlation with a (random effect) variance parameter. Through standard equations, we will show now that the within-cluster correlation is restricted to be positive, since the random effect variance cannot be negative. The one-way random effects model is most commonly used for describing continuous data that are clustered in one way. Without making an explicit distinction between a random variable and a realized value, the outcome $y_{ij}$ is the $j$-th observation in the $i$-th cluster and expressed as the sum of the general mean, $\mu$, random effect $\alpha_{i}$,  and residual error $e_{ij}$,
\begin{eqnarray}\label{simple_conditional_model}
y_{ij} &=& \mu + \alpha_{i} + e_{ij}, \\
\alpha_{i} &\sim& N(0,\tau), \nonumber \\
e_{ij} &\sim& N(0,\sigma^2). \nonumber
\end{eqnarray}
The within-cluster error variance, $\sigma^2$, represents the variation in observations in each cluster $i$ given the random effect $\alpha_i$. Aside from the term random effect, the $\alpha_i$ is also referred to as the \textit{blocking} factor, \textit{grouping} factor, or the \textit{treatment} factor. The random effect is assumed to be normally distributed with mean zero and variance $\tau$. The random effect variance $\tau$ represents the variation in random intercepts across clusters, and is often referred to as the between-cluster variance. Indeed, this restricts the $\tau$ to positive values only. The BCSM approach will relax this restriction by introducing a different representation of the model, and will later show why this is relevant.

The random effect variance parameter is not presented in a squared notation, since this variance parameter also represents the covariance among clustered observations. A covariance parameter is not restricted to be positive, but squared terms always are. The relation between the covariance and $\tau$ becomes immediately apparent when considering the covariance between the two clustered observations $j$ and $l$ in group $i$, which is represented by
\begin{eqnarray}\label{covariance_i_l}
Cov\left(y_{ij},y_{il} \right) &=& Cov\left(E\left(y_{ij}\mid \alpha_i\right),E\left(y_{il} \mid \alpha_i \right)\right) + E\left(Cov\left(y_{ij},y_{il} \mid \alpha_i \right)\right), \nonumber \\
& = & Cov\left(\mu + \alpha_{i},\mu + \alpha_{i}\right) + 0, \nonumber \\
& = &  Cov\left(\alpha_{i}, \alpha_{i}\right) = Var\left(\alpha_i \right) = \tau.
\end{eqnarray}
Then, the variance of an observation equals
\begin{eqnarray}
Var\left(y_{ij} \right) & = & Var\left(E\left(y_{ij}\mid e_{ij}\right)\right) + E\left(Var\left(y_{ij} \mid \alpha_i \right)\right), \nonumber \\
& = & \sigma^2 + \tau.\label{varianceY}
\end{eqnarray}
{\color{black}
The covariance structure represented in Equation (\ref{covariance_i_l}) and (\ref{varianceY}) has an additive form, which is used later on in the construction of the posterior distribution of the covariance parameters. This additive form is not limited to independently distributed level-1 residuals. In Section A of the Supplementary Materials it is shown that the additive form of the covariance structure is retained for correlated level-1 residuals.
}

The intra-class correlation (ICC) represents the proportion of variance explained by the clustering. This is represented by
\begin{equation}\label{ICC}
\rho = \frac{Cov(y_{ij}, y_{il})}{Var(y_{ij})} = \frac{\tau}{\sigma^{2} +\tau}.
\end{equation}
Under standard MLM, the ICC is restricted to be positive since both the numerator $(\tau)$ and the denominator $(\sigma^{2}+\tau)$ are variance parameters. Generally, the interpretation of $\rho$ stops here, as the general tendency is to think that an ICC cannot be negative, restricting $\rho$ to lie between zero and one \parencite[see for example][]{Eldridge2009, Huang2018}.

However, the covariance component in the numerator in Equation \eqref{ICC} could also be negative if $\tau$ represents the covariance among clustered observations, and not also the random effect variance. It is this double function of the random effect variance parameter $\tau$ that restricts the covariance among clustered observations, and the ICC, to be positive.

\section{Examples of Negative Clustering}
\begin{figure}[hbt!]
	\centering
	\includegraphics[width=150mm]{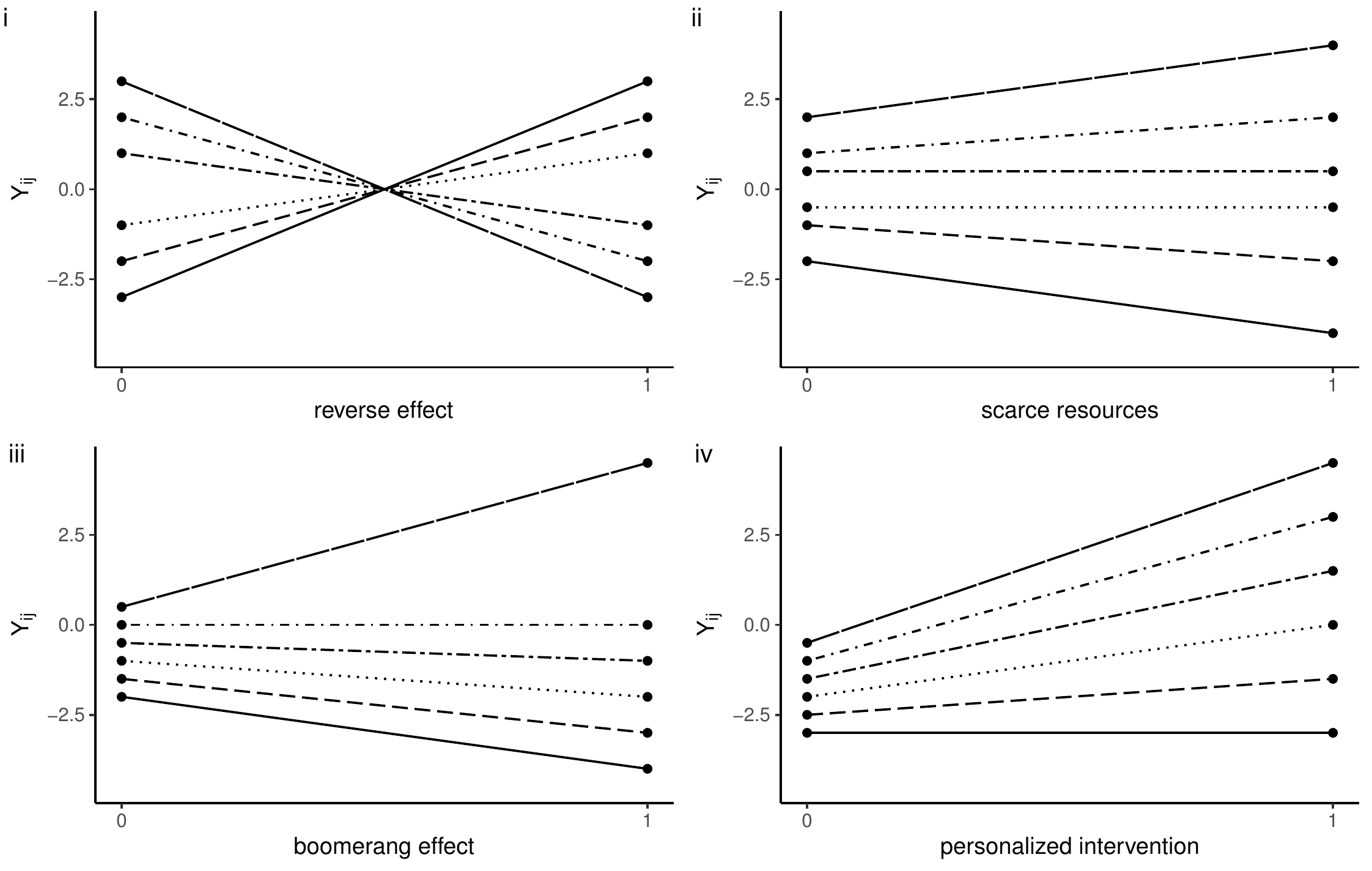}
	\caption{Four examples where the cluster variance $\tau$ is negative (scenario \Rn{1} - \Rn{4}). Outcome $Y_{ij}$, observed two times (e.g., repeated measures, 0 for the pre-, and 1 for the post-observation). Scenario \Rn{1} shows a reverse effect, all pre-observations are reversed (`reverse effect') at the post-measurement; \Rn{2} shows a setting where individuals are competing for scarce resources (e.g., one's pain is the other's gain); \Rn{3} shows how other group members decrease as a reaction to the increase of another group member \parencite[`boomerang effect';][]{Kenny2002}; and \Rn{4} shows that clustered individuals have dissimilar trajectories related to the personalized intervention.}
	\label{fig:BCSM_introduction_negative_associations}
\end{figure}

We discuss several examples where researchers either encountered negative ICC values, or where they could be expected. Note that it is currently difficult to give a literature overview: researchers do not report on negative ICCs, nor that it is well-known that these values in fact occur, and the common statistical software packages do not allow for negative associations between clustered observations. We visualize depict the following examples in Figure \ref{fig:BCSM_introduction_negative_associations}.

\subsubsection{Multidisciplinarity}
Nowadays in science --but also in society and everywhere where people collaborate-- it is increasingly important to think, act and create across boundaries. Therefore, the cultural, ethical or scientific background of one individual should differ from that of the others, which stimulates the dissimilarity in a group. This is a pattern that can also be seen in the police force, in politics, and education. Other forms of diversity in a group are co-morbidity, the number of co-morbid diseases increases with age, or through smoking, development of dementia and other diseases, human migration.

\subsubsection{The boomerang effect}
Another factor that can create negative clustering dependencies is referred to as the `boomerang effect' \parencite{Kenny2002}: one set of observations may influence the other observations in the cluster to be different. Figure \ref{fig:BCSM_introduction_negative_associations} gives an illustration of such pattern: in a pre-test post-test (or repeated measurements) design, rather than individuals behaving more similar, one (or a small set of observations) increases dissimilarity in a cluster. For example, in group or family counselling, the behaviour of a narcissistic individual might decrease the self-worth of others. A similar pattern might be noticed with a pessimistic individual: one pessimistic individual may increase the mood of others. In Figure \ref{fig:BCSM_introduction_negative_associations}, this boomerang effect is shown for a repeated measures setting with two measurement occasions, where the dissimilarity is increasing in the cluster.

\subsubsection{Competing}
Another source that can cause negative clustering effects is competing, which was suggested by \textcite{Pryseley2011}. Figure \ref{fig:BCSM_introduction_negative_associations}, an illustration is given of how competing can increase the variances among the mebers of a group. Individuals often compete for the allocation of scare resources within the same group. The examples suggested by \textcite{Pryseley2011} are litter mates, division of a fixed reward, speaking time, and leadership. In Figure \ref{fig:BCSM_introduction_negative_associations}, this phenomena is referred to as \textit{one's pain is another's gain}.

\subsubsection{Personalized interventions}
The typical situation in studies of psychotherapy process and outcome is that one counsellor treats several clients \parencite{Baldwin2013b}. When the clients who see the same counsellor are more similar to each other than those clients who are treated by different counsellors, outcomes of clients with the same counsellor are expected to be positively correlated. The counsellor treats clients in a similar way, which leads to a common positive correlation among the treated clients. Although the efficacy -- or clustering effect -- of the counsellor is well-known to be important, it is not always assessed. Doing so is straightforward in the MLM approach \parencites{Raudenbush2001, Baldwin2013b, Kenny2009, Marcus2009}. However, when the counsellor provides a personalized treatment, the effects of each treatment can differ substantially across clients. Personalized interventions are designed for the individual \parencite{Smink2019_Frontiers}: \textit{what} treatment, by \textit{whom}, is most effective for \textit{this} individual with \textit{that} specific problem, and under \textit{which} set of \textit{circumstances} \parencite[][p. 111]{Paul1967}? As a result, dissimilarity in a counsellor's client group can occur when for some individuals the personalized treatment works well but not for others. This can lead to a negative correlation among the treated clients of a counsellor. In fact, a negative correlation would indicate that some clients benefit highly from the personalized treatment, where for others positive treatment effects are more difficult to realize. The negative correlations also provide information about the counsellor who is able to improve the treatment of clients through personalization leading to dissimilar client results, since clients still respond in different ways to a personalized treatment.

\section{The Bayesian Covariance Structure Model}

The general idea of BCSM is to model directly the dependence structure of the data, and not indirectly through random effect parameters. This dependence structure can be implied by random effects. The BCSM is a more general approach for clustered data, since it can also identify a negative dependence structure and a dependence structure implied by non-identifiable random effects. BCSMs have been developed for different applications to deal with complex correlated data structures \parencite{Fox2017,Klotzke2019a, Klotzke2019b, Mulder2019}.

Consider the error terms $\alpha_i$ and $\mathbf{e}_{i}=(e_{i1},\ldots,e_{in})$ to describe the dependence structure for the clustered observations. The error component for cluster $i$, $\mathbf{E}_i = \alpha_i + \mathbf{e}_{i}$, is assumed to be multivariate normally distributed, where the covariance matrix comprehends the common covariance among the clustered observations (Equation \ref{covariance_i_l}) on the non-diagonal and the total variance (Equation \ref{varianceY}) on the diagonal. It follows that,
\begin{eqnarray}\label{BCSM1}
\mathbf{y}_i & = & \mu + \mathbf{E}_i,  \\
 \mathbf{E}_i  & \sim & N(0,\bm{\Sigma}), \nonumber
\end{eqnarray}
where
\begin{equation}\label{variance_covariance_matrix}
\mathbf{\Sigma} =
\begin{bmatrix}
\sigma_{}^{2} + \tau & \tau                  & \dots   & \tau                  \\
\tau                  & \sigma_{}^{2} + \tau & \dots   & \vdots                \\
\vdots                & \vdots                & \ddots  & \vdots                \\
\tau                  & \dots                 & \tau    & \sigma_{}^{2} + \tau
\end{bmatrix}.
\end{equation}
Under the BCSM, parameter $\tau$ is no longer a variance parameter and only represents the common covariance among clustered observations. The $\tau$ is a covariance parameter and not a variance parameter. This has three important implications: 1) $\tau$ can now also be negative, 2) zero is no longer the boundary value for $\tau$, and 3) $\tau$ is not estimated as the random intercept variance. Indeed, negative values for $\tau$ are now perfectly acceptable, since this merely corresponds to the occurrence of \textit{negative} within-cluster correlation. The implications of negative clustering effects will be discussed later. The only requirement is that the covariance matrix is positive definite, which is the case when $\tau  > - \sigma^{2}/n$ (which will be shown later).

{\color{black}
The lower bound for $\tau$ implies that the correlation between two clustered observations is more than $(-\sigma^2/n)/(\sigma^2-\sigma^2/n)=-1/(n-1)$, which goes to zero when increasing the cluster size $n$. This is not really an issue. First, in general it is simply not possible to have a common negative correlation among many observations, so the BCSM cannot capture this. Second, there are many multilevel applications for small cluster sizes (e.g. family and twin studies, professional teams, repeated measurements). Third, our main motivation for modeling negative correlation is to identify individual variation in the cluster effect, which manifests itself by a negative correlation among a (small) group of individuals. When a cluster effect is beneficial for some it is not for others in the same cluster due to a negative within-cluster correlation. When the cluster size becomes too large, it is no longer possible to make this distinction and to identify individualized effects.
}

\subsubsection{Type of Dependence}
It is straightforward to represent the covariance matrix in matrix notation. Assume that each cluster $i$ has $n$ observations, then $\bm{\Sigma}=\sigma^2 \mathbf{I}_n + \tau\mathbf{J}_n$, where the $\mathbf{J}_n$ is a matrix of dimension $n$ with all elements equal to one and $\mathbf{I}_n$ is the identity matrix of dimension $n$. The dependence structure of this covariance matrix $\bm{\Sigma}$ is straightforward: if there is no clustering in the data, the covariance $\tau$ is not present (e.g. $\tau = 0$). If $\tau$ is positive, the observations are assumed to be positively correlated and the dependence structure is similar to that of the random intercept model in Equation \eqref{simple_conditional_model}. If $\tau$ is negative, the observations are negatively correlated within a cluster, a dependence structure that cannot be represented by a random intercept model. Thus, the BCSM elegantly represents three nested models depending only on the sign and value of the covariance parameter. Indeed, the BCSM simply extends the range of possible values to include zero and negative values, without changing the interpretation of positive values.

\subsubsection{Multiple Types of Dependence}
In our real data example, for each client a pre-intervention and post-intervention score was observed, and clients were treated by counselors. Thus, observations were clustered by clients (type A clustering), who were again clustered by counselors (type B clustering). The BCSM can be extended to describe any additional type of clustering. To illustrate this, we consider our real-data design, where observations were clustered according to type A, as described in Equation \eqref{simple_conditional_model}, and that those clustered observations are again clustered according to type B. In the two-way random effects model, a random effect $\beta_{(i)j}$ can be introduced that represents the clustering of observations according to type B, which is represented by
\begin{eqnarray}\label{twowayconditional}
y_{ijk} &=& \mu + \alpha_i + \beta_{(i)j} + e_{ijk}, \\
\alpha_{i} &\sim& N(0,\tau_a), \nonumber \\
\beta_{(i)j} &\sim& N(0,\tau_b), \nonumber \\
e_{ijk} &\sim& N(0,\sigma^2). \nonumber
\end{eqnarray}
It follows that objects in cluster $i$ are nested (type A), where the $\tau_a$ represents the common dependence among the clustered observations. Within each cluster $i$, observations in each cluster $ij$ are again nested (type B), where the $\tau_b$ represents the dependence among those clustered observations. The random effect variance parameters $\tau_a$ and the $\tau_b$ represent the dependence among clustered observations but are both restricted to be positive.

In the BCSM for this two-way (nested) structure, the dependence structure is directly modelled. To be complete, the covariance matrix is given for this design. Let $b$ clusters of type B each of size $n$ be nested within the cluster of type A, with in total $a$ type-A clusters. Then, the BCSM covariance matrix is represented by
\begin{eqnarray}\label{twowaySigma}
\bm{\Sigma} & = & \left(\mathbf{I}_{nb}\sigma^2 + \mathbf{J}_{nb}\tau_a\right) + \left(\mathbf{I}_b\otimes\mathbf{J}_n\right)\tau_b.
\end{eqnarray}
The covariance matrix of the one-way clustering is extended with an extra component that displays the nesting of observations in type B clusters. The Kronecker product $\otimes$ is needed to define which of the observations in each cluster A are again nested according to cluster B. It states that the $b$ blocks of $n$ observations are clustered with a common dependence of $\tau_b$. The BCSM for the two-way clustered data is represented by Equation \eqref{BCSM1} with the covariance matrix defined in Equation \eqref{twowaySigma}. {\color{black} The covariance matrix needs to be positive definite, which leads to the restriction $\tau_b > -\sigma^2/n$ and $\tau_a > -(\tau_b/b + \sigma^2/(bn))$. The restrictions follow from the derivation of the posterior distributions for $\tau_a$ and $\tau_b$, which is shown later. More formally, it is shown in the Supplementary Materials (Section B) that the restrictions also follow from the expression for the determinant.
}

In the BCSM any type of clustering is directly modelled through the structured covariance matrix, and this covariance matrix can represent multi-way structured data. Furthermore, a hybrid version is also possible, where the mean component also includes random effect parameters. For instance, a hybrid version of a two-way BCSM can be defined by including the random effect $\beta_{(i)j}$ in the mean term with the structured covariance matrix of the one-way model in Equation \eqref{variance_covariance_matrix}. The BCSM represented in Equation \eqref{BCSM1} is also easily extended to include explanatory variables with fixed effects. Let $\mu=\mathbf{X}_i \bm{\beta}_f$, the (design) matrix $\mathbf{X}_i$ contains the explanatory variables for cluster $i$ and the $\bm{\beta}_f$ represents the regression effects of the variables.

\section{Advantages of BCSM over MLM}
To summarize the previous section: the BCSM is a novel Bayesian modelling framework in which the covariance structure of a (complex) dependence structure is directly modelled. This makes the BCSM more flexible and more general than standard MLMs. We give an overview of the specific features of BCSM in comparison to MLM. Next to a theoretical discussion of the advantages of BCSM, we specifically designed our simulation and real-data study to provide more evidence in support of these claims.

\subsubsection{Modelling Negative Clustering Effects}
Negative correlations among clustered observations cannot be modelled with the MLM. In the MLM, a positive correlation is modelled through a shared group-specific effect among the group members. This modelling concept cannot be translated to model negative dependences, since sharing a common component always leads to a positive association. The BCSM has been developed with the purpose to model in a similar way positive as well as negative correlations among clustered observations, while using a common dependence structure across groups.

Although a well-known and widely applicable statistical model for negatively correlated clustered data is lacking, the negative effects of ignoring negatively correlated clustered data has been mentioned in the literature. Ignoring a positive correlation in the data leads to an increase of the Type-I error, $p$-values that are biased downwards and confidence intervals that are too narrow. Standard errors of fixed regression effects are smaller than they should be, feigning a precision of the estimates that is not actually supported in practice, leading to spurious and erroneous results of statistical significance \parencite{Kenny1998}. For relatively small clustering effects (i.e. for small values of the ICC),  \textcite{Barcikowski1981} showed that for instance an ICC of 0.05 and 100 observations per group already inflates the probability of a Type-I error to 0.43. Next to ignoring a positive correlation, the ignorance of a (small) negative correlation within groups leads to opposite effects compared to ignoring positive correlation within groups: a deflation of Type-I errors, $p$-values that are biased upwards and overestimated SEs (i.e. confidence intervals that are too wide). This deflation of the Type-I error, when ignoring a negative correlation has been mentioned by other researchers \parencites{Barcikowski1981, Rosner1999}. \textcite{Nielsen2021} also quantified in detail the negative effects of ignoring the negative correlation.

There is an apparent risk of ignoring dissimilarity (i.e. negative correlations) within clusters. Even the smallest dissimilarity between clustered observations can seriously inflate the probability of a Type-I error. \textcite{Kenny2002} argue along the same lines: if \textit{positive} clustering effects can cause various statistical problems, then so do \textit{negative} clustering effects. It is well known that when the ICC is greater than zero, which often occur in psychology \parencite{Hox2010, Hoyle2001}, the use of MLMs is advised to analyse the data. However, when the clustered data are negatively correlated, the dissimilarity in the clustered data is usually ignored, despite the negative effects of ignoring a negative ICC. Even though others --such as \textcite{Kenny2002} and \textcite{Pryseley2011}-- already drew attention to this phenomenon of dissimilarity, it is obvious that the negative counterpart is less well understood. Furthermore, a more pragmatic reason is that until recently, the tools to study negatively correlated data is lacking \parencite[although there are of course exceptions,][]{Verbeke2003, Molenberghs2011}. The opinion is that this risk of ignoring a non-zero ICC is currently even greater under negative clustering effects, as MLMs cannot assess negative clustering effects, and the effects of negative clustering effects appear to be less well-known by researchers.

\subsubsection{Go Beyond Sample Size Restrictions}
To obtain stable parameter estimates for the MLM, the sample size needs to be sufficient for the different levels of the model. For the one-way random effects model, a sufficient number of clusters is needed to estimate the variability across groups. For a multi-way random effects model, for each clustering type a sufficient number of clusters are needed to obtain a stable random effect variance estimate. \textcite{Maas2005} reported that a small sample size at level two can lead to biased estimates of the second-level standard errors. A small number of level-two groups can lead to a zero level-two variance estimate, indicating that there is simply not enough information. The Bayesian approach can introduce a prior to by-pass this problem. However, a prior distribution can force the variance estimate to be positive. This can highly depend on the specified prior and might not represent correctly the variation across clusters in the population. Furthermore, the motivation for doing a multilevel analysis is that the sample size within each cluster is less than overwhelming. Then, the cluster-level variance is used as a weight to reduce the error in the cluster-specific estimates by pooling information across clusters. However, the shrinkage in the cluster-specific estimates might be less than desired, when the cluster-level variance is overestimated.

In the BCSM, the dependence structure is modelled through a common covariance parameter for the clustered observations. This reduces the sample size restrictions for the BCSM compared to the MLM. Furthermore, a prior for a covariance parameter is not restricted to positive values. The BCSM can be applied to a two-stage (or multi-stage) sample, where clusters are sampled independently, and subsequently observations within each cluster are independently sampled. However, by modelling directly the covariance among clustered observations, the BCSM also applies to a stratified sample in which independent samples are drawn for the considered clusters.

We will demonstrate in our simulation study that even for two clusters stable covariance parameter estimates can be obtained. Furthermore, the BCSM will prove to be very useful for analysing small data sets. Under the BCSM, the practical definition of what is considered a small sample size changes considerably. The BCSM in Equation \eqref{BCSM1} does not contain any cluster-specific parameters, although cluster-specific estimates can be obtained from fitted residuals. As a result, cluster-level variance estimates are not needed to shrink cluster-specific parameters. This avoids the issue of estimating the variability across clusters, and to use those estimates to reduce errors in the clusters-specific parameter estimates by shrinking them. This makes the BCSM much more suitable for small sample sizes than the MLM. In the BCSM, it is not needed to explicitly model variability across clusters and to estimate any cluster-specific (i.e. random effect) parameters. Furthermore, due to the Bayesian modelling approach, it is also not necessary to rely on large sample theory to make statistical inferences.

\subsubsection{Model Complexity}
The BCSM represents a far more parsimonious way to model a dependence structure than the random effects approach in MLM. Under the BCSM, the number of covariance parameters to model the dependence structure does not depend on the sample size. This in contrast to the MLM, where the required number of random effect parameters depends on the number of clusters. Indeed, increasing the number of clusters does not affect the complexity of the BCSM, where the MLM becomes more complex. Furthermore, for each additional type of clustering, the dimensionality of the MLM increases and requires an additional set of random effect parameters, where the BCSM requires just one additional covariance parameter.

The BCSM can even model a dependence structure implied by non-identified random effects. For instance, assume pre-intervention and post-intervention data of persons, and let $\alpha_{i}$ denote the person-specific random effect for the post-measurement of person $i$, which is normally distributed with variance $\tau_a$. The cluster size is $n=1$ (i.e. each person has one post-measurement), which makes it impossible to estimate the random effect $\alpha_{i}$ and the variance $\tau_a$. Under the BCSM, the parameter $\tau_a$ is identified and can be estimated, which provides information about the dependence of the post-intervention measurements. The BCSM approach is straightforward and elegant: the covariance matrix has a common error variance $\sigma^2$ for the pre-intervention measurements and a variance component $\tau_a$ is added to the common error variance for the post-intervention measurements. The heteroscedastic error variances of the covariance matrix are identified and can be motivated by the (unidentified) random effect $\alpha_{i}$. Thus, under the BCSM, the dependence structure of a random interaction effect can be estimated from clusters which only have one observation.

\subsubsection{Unbiased Estimator: Include the Entire Parameter Space}
Common maximum likelihood (ML) and Bayesian estimation methods restrict the random effect variance estimate to be positive. Bayesian methods use a prior which assigns a positive density to non-negative values; ML methods usually restrict the variance estimate to be positive, although negative variance estimates are possible (see below). This leads to biased parameter estimates. We show here that the random intercept model, Equation \eqref{simple_conditional_model}, gives support to data sets for which the ML estimate is negative. As a result, when not allowing negative variance estimates, the ML estimator is biased, since the negative parameter space of the sampling distribution of the estimator is ignored. This also holds for the restricted maximum likelihood estimator and for (un)balanced designs. {\color{black} In the BCSM, the prior for the covariance parameter includes the negative parameter space, for all values for which the covariance matrix is positive definite. The Bayesian estimator under the BCSM is not unbiased from a sampling theory approach, however the entire parameter space is taken into account and a uniform prior can be specified that does not favor any value above another.
}

A negative ML estimate of the random effect variance has received attention \parencite{Kenny2002,  Molenberghs2007, Molenberghs2011, Pryseley2011, Oliveira2017, Verbeke2003, ElLeithy2016, Loeys2013}, partly due to the embarrassment of obtaining a negative estimate for a parameter which by definition is non-negative \parencite[][p.60]{Searle1992}. For the random intercept model in Equation \eqref{simple_conditional_model}, it can be easily seen that the ML estimate for the random intercept variance $\tau$ can be negative depending on the observed between-cluster and (within-cluster) error sum of squares. For balanced groups, the two sums of squares are considered to estimate the covariance component $\tau$,
\begin{eqnarray}
SS_{A} & = &\sum_{i=1}^{a} n \left(\overline{y}_i - \overline{y} \right)^2, \nonumber \\
SS_{E} &= & \sum_{i=1}^{a}  \sum_{j=1}^{n} \left(y_{ij} - \overline{y}_j \right)^2. \nonumber
\end{eqnarray}
The sum of squares $SS_{A}/a$ has expected value $n\tau+\sigma^2$. It follows that,
\begin{eqnarray}
\hat{\tau} & = & \frac{\frac{SS_{A}}{a} - \frac{SS_{E}}{n(a-1)}}{n} \nonumber \\
                    & = & \frac{\frac{SS_{A}}{a} - MSE}{n}, \nonumber
\end{eqnarray}
using the $MSE$ as an estimator for $\sigma^2$. The estimate for $\tau$ is negative when $MSE > SS_{A}/a$. The negative estimates are neglected or referred to as statistically incorrect, restricting $\tau$ to be positive, $0< \tau \le \infty$. However, the ML estimate is not necessarily in this parameter space, which occurs with probability $P(MSE > SS_{A}/a)$. As described by \textcite{McCulloch2008}, the ML estimator has two possible outcomes
\begin{eqnarray}
\hat{\tau} & = & \left\{
                     \begin{array}{lcc}
                       \hat{\tau} & \text{if} & SS_{A}/a \ge MSE \\
                       0 & \text{if} & SS_{A}/a < MSE.
                     \end{array}
                     \right.\nonumber
\end{eqnarray}
The estimate of the variance is restricted to be zero, when the data gives support to a negative estimate. Of course this makes sense, since $\tau$  represents a variance component. However, for $\tau<0$ there is cluster dissimilarity, which will be interpreted incorrectly as cluster similarity when $\tau$ is restricted to be positive.

\subsubsection{Solving Boundary Issues}

In the MLM, the random effect variance is restricted to be greater or equal to zero. This value of zero is a lower bound but also of specific interest. A random effect variance of zero implies that the groups do not differ, where a positive variance implies that the groups differ. It is well-known that classical test procedures such as the likelihood-ratio test can break down and leads to inconsistent testing, when testing if a parameter lies on the boundary of the parameter space.

In the Bayesian framework, test and estimation methods depend on the specified prior distributions. Specifying a prior for a random effect variance is a complicated task, since the point zero is a boundary value. The popular conjugate inverse-gamma prior only gives support to positive values. The exact specification of the prior depends on the hyper parameter values. When the variance is near zero the hyper parameters need to be close to zero. Most often the default inverse-gamma prior is sharply peaked near zero to give support to variance values near zero. Thus, an objective (non-informative) prior specification is not possible without knowing the true parameter value. Otherwise stated, the posterior distribution is sensitive to the hyper parameter values of the inverse-gamma distribution. \textcite{Gelman2006a} recommended different classes of priors such as the half-$t$ family of prior distributions, to improve the behaviour of the prior near zero. However, the priors are not completely objective and, in general, place too much mass on higher variance values when the true value is close to zero. This phenomenon is shown in our simulation study.

Under the BCSM, the value $\tau=0$ is not a lower bound. Therefore, a noninformative prior can be specified for those parameter values that ensure a positive-definite covariance matrix. Following \textcites{Fox2017}, a truncated shifted inverse-gamma prior can be specified that allows the parameter space to cover also negative values while enforcing sufficient rules for the positive definiteness of the covariance matrix. These priors are not sharply peaked near zero such as the default inverse-gamma priors but remain uninformative about the presence of negative, positive, or zero correlation. In addition, with the shifted inverse-gamma prior, more accurate estimates of a very small random-effect variance can be obtained by avoiding too much prior support for higher parameter values.

\section{Parameter Estimation for the BCSM}

A general technique is proposed to estimate the model parameters of the BCSM. The estimation method is based on a Gibbs sampler (Markov chain Monte Carlo, MCMC), where the variance components of the BCSM can be directly sampled from their conditional posterior distributions. The posterior distribution of each variance component can be analytically derived from which parameter values can be directly sampled. This technique is based on a balanced design, which means that the number of  observations is equal across the same type of clustering. Although the BCSM is by no means limited to balanced designs alone, the extension to unbalanced designs is beyond the scope of our current study.

\subsubsection{One-way Classification}
Three steps can be defined to construct the MCMC algorithm for the BCSM for the one-way classification in Equation \eqref{BCSM1}. In step 1, the expected within-sum of squares ($SS_E$) is derived to construct the posterior distribution of the variance parameter $\sigma^2$. In a similar method, in step 2, the expected between-sum of squares ($SS_A$) is derived. In step 3, a shift-parameter is introduced for the result of step 2, to obtain the posterior distribution of the covariance parameter $\tau$.

The posterior distributions of the variance components $\sigma^2$ and $\tau$ are derived. In this model, the total sum of squares $(SS_{T})$ is partitioned in a between- and within-sum of squares, referred to as $SS_{E}$ and $SS_{A}$ (type-A clustering), respectively,
\begin{eqnarray}
SS_{T} & = & SS_{A} + SS_{E} \\
\sum_{i=1}^{a} \sum_{j=1}^{n} \left(y_{ij}-\overline{y}_{..} \right)^2 & = & \sum_{i=1}^{a} n \left(\overline{y}_{i.} - \overline{y}_{..}\right)^2 +
\sum_{i=1}^{a} \sum_{j=1}^{n} \left(y_{ij} - \overline{y}_{i.}\right)^2,  \nonumber
\end{eqnarray}
where $\overline{y}_{..} = \sum_{i=1}^{a}\sum_{j=1}^{n} y_{ij}/(na)$ and $\overline{y}_{i.} = \sum_{j=1}^{n} y_{ij}/n$. The part of the likelihood that includes the general mean is excluded. This follows from partitioning the likelihood;
\begin{eqnarray}
p\left(\mathbf{y} \mid \mu,\sigma^2,\tau \right) = p\left(\mu \mid \overline{y}_{..}\right)p\left(\sigma^2,\tau \mid SS_{E}, SS_{A}\right), \nonumber
\end{eqnarray}
see, for instance, \textcite{McCulloch2008}. As they follow directly from standard Bayesian linear regression theory \parencite{Gelman2013}, the posterior distributions of fixed effect parameters are not discussed.

The conditional model in Equation \eqref{simple_conditional_model} in which observations are conditionally independently distributed given cluster-specific parameters, is used to find the model expressions for the cluster and sample-averaged observations. It follows that
\begin{eqnarray}
\overline{y}_{i.} &=& \mu + \alpha_{i} + \overline{e}_{i.} \label{meanyi} \\
\overline{y}_{..} &=& \mu + \overline{\alpha}_{.} + \overline{e}_{..}, \label{meany}
\end{eqnarray}
where $\overline{e}_{i.} \sim N(0,\sigma^2/n)$ and $\overline{e}_{..} \sim N(0,\sigma^2/(na))$. The expressions are used to obtain the expected sum of squares under the model.

\textbf{Step 1} is carried out. Therefore, the expected value of the $SS_{E}$ is derived by integrating the model expression for the cluster mean (Equation \ref{meanyi}):
\allowdisplaybreaks
\begin{eqnarray}
E(SS_{E}) & = & E\left(\sum_{i = 1}^{a}\sum_{j = 1}^{n} \left(y_{ij} - \overline{y}_{i.}\right)^{2}\right) \nonumber \\
&= & \sum_{i = 1}^{a}\sum_{j = 1}^{n} E\left((\mu + \alpha_{i} + e_{ij}) - (\mu + \alpha_{i} + \overline{e}_{i.})\right)^{2}
\nonumber \\
&= & \sum_{i = 1}^{a}\sum_{j = 1}^{n} E\left(e_{ij} - \overline{e}_{i.} \right)^{2} \nonumber \\
& = & \sum_{i = 1}^{a}\sum_{j = 1}^{n}  E\left(e^2_{ij}\right) - E\left(\overline{e}^2_{i.}\right) \nonumber \\
& = & an\left(\sigma^{2} - \frac{\sigma^{2}}{n}\right) \nonumber \\
&= & a(n - 1)\sigma^{2},
\end{eqnarray}
where in the extraction of the binomial product the inner product cancels (from the second to the third expression, and the third to the fourth expression), since the expected value of each error term is zero. For a balanced design and pairwise independent $SS_{E}$ components, the $SS_{E}$ divided by their expected value is (central) chi-square distributed \parencite[][p.174]{Searle1971}.

Assume an inverse-gamma prior for $\sigma^2$, $\sigma^2\sim IG(g_1/2,g_2/2)$. Then, the posterior distribution of the $\sigma^2$ is an inverse-gamma distribution with $SS_{E}$ as the sufficient statistic \parencite{Gelman2013},
\begin{eqnarray}
p\left(\sigma^2 \mid \mathbf{y} \right) \propto \left(\sigma^2\right)^{-((a(n-1)+g1)/2+1)}\exp\left(-\frac{(SS_{E}+g_2)/2}{\sigma^2}\right)
\end{eqnarray}
with shape parameter $(g_1+a(n-1))/2$  and scale parameter $(SS_{E}+g_2)/2$. For $g_1=0$ and $g_2=0$ the uninformative reference prior is specified for $\sigma^2$.

In \textbf{step 2}, a similar procedure is followed for the covariance parameter $\tau$. Consider the between sum of squares $SS_{A}$,
\allowdisplaybreaks
\begin{eqnarray}
E(SS_{A}) & = & E\left(n\sum_{i = 1}^{a}{\left( \overline{y}_{i.} - \overline{y}_{..} \right)}^{2}\right) \nonumber \\
&= & n\sum_{i=1}^{a}E\left((\mu + \alpha_{i} + \overline{e}_{i.}) - (\mu + \overline{\alpha}_{.} + \overline{e}_{..}) \right)^{2} \nonumber \\
&= & n\sum_{i=1}^{a} E\left((\alpha_{i}-\overline{\alpha}_{.}) + (\overline{e}_{i.} - \overline{e}_{..})\right)^{2}  \nonumber \\
&= & n\sum_{i=1}^{a} E\left(\alpha_{i}-\overline{\alpha}_{.}\right)^2 + E\left(\overline{e}_{i.} - \overline{e}_{..}\right)^{2}  \nonumber \\
& = & n\sum_{i=1}^{a} \left(E(\alpha^2_{i}) - E(\overline{\alpha}^2_{.})\right) + \left(E\left(\overline{e}^2_{i.}\right) - E\left(\overline{e}^2_{..}\right)\right) \nonumber \\
&= & an\left(\left(\tau - \frac{\tau}{a}\right) + \left(\frac{\sigma^2}{n} - \frac{\sigma^2}{an} \right)\right)  \nonumber \\
& = & (a-1)\left(n\tau + \sigma^2\right),
\end{eqnarray}
where the inner product of the binomial products is again zero, since the expected error terms are equal to zero. The $SS_{A}/n$ is considered as the sufficient statistic for the term $\lambda = \tau + \sigma^{2}/n$, which has an inverse-gamma distribution. The $\lambda$ is restricted to be positive, which means that $\tau >  -\sigma^{2}/n$ with $\sigma^{2}>0$.

In \textbf{step 3}, the shift parameter is introduced, which is the term $\sigma^2/n$, and allows the $\tau$ to take on negative values. This restriction on the parameter space of $\tau$ can be defined in the noninformative prior for $\tau$;
\begin{eqnarray}
p\left(\tau \mid \sigma^2 \right) \propto \left(\tau + \sigma^{2}/n\right)^{-1},
\end{eqnarray}
since it restricts the $\tau$ to be greater than $-\sigma^{2}/n$ with $\lambda = \tau + \sigma^{2}/n$ restricted to be greater than zero. Following \textcite{Fox2017} and \textcite{Klotzke2019a}, the posterior distribution of $\tau$ is referred to as a shifted inverse-gamma distribution
\begin{eqnarray}
p\left(\tau \mid \mathbf{y},\sigma^2 \right) \propto \left(\tau + \sigma^2/n\right)^{-((a-1)/2+1)}\exp\left(-\frac{(SS_{A}/n)/2}{\tau+\sigma^2/n}\right). \nonumber
\end{eqnarray}
It can also be shown that for all $\tau$ values above this lower bound the covariance matrix in Equation \eqref{covariance_i_l} is positive definite \parencite{Fox2017}. Parameter values from this shifted inverse gamma distribution can be obtained by sampling $\lambda^{(m)}$ from an inverse-gamma distribution with $(a-1)/2$ degrees of freedom and scale parameter $(SS_{A}/n)/2$ in iteration $m$. Then, a sampled value for $\tau$ is obtained by subtracting the sampled value for $\sigma^{2}$, $(\lambda^{(m)} - \sigma^{2}/n)$.

\subsubsection{Two-way Classification}
This procedure to derive the posterior distributions of the variance components can be extended to covariance parameters for other cross-classified and/or nested factors. Without giving a general description, the two-way nested classification model in Equation \eqref{twowayconditional} is considered to illustrate the procedure for two types of clustering (referred to as type A and type B). Again three steps can be defined, where step 1 is similar to the step 1 for the one-way classification. Then, step 2a (obtain expected between sum of squares) and 3a (derive shift parameter) are defined to obtain the posterior distribution of parameter $\tau_a$ for the clustering of type A. Analogously, step 2b and 3b are defined for the $\tau_b$ for the clustering of type B.

The total sum of squares is partitioned in three components, the total sum of squares $(SS_{T})$, a sum of squares $SS_{A}$ (cluster A), a sum of squares $SS_{B}$ (cluster B) and a within-sum of squares $(SS_{E})$:
\begin{eqnarray}
SS_{T} & = & SS_{A} + SS_{B} + SS_{e} \\
\sum_{i=1}^{a} \sum_{j=1}^{b}\sum_{k=1}^{n} \left(y_{ijk}-\overline{y}_{...} \right)^2 & = & \sum_{i=1}^{a} nb \left(\overline{y}_{i..} - \overline{y}_{...}\right)^2 + \sum_{i=1}^{a} \sum_{j=1}^{b} n\left(\overline{y}_{ij.} - \overline{y}_{i.}\right)^2  \nonumber \\
&&+\sum_{i=1}^{a} \sum_{j=1}^{b} \sum_{k=1}^{n} \left(y_{ijk} - \overline{y}_{ij.}\right)^2.  \nonumber
\end{eqnarray}
\textbf{Step 1}: the expected value of the $SS_{E}$ is derived,
\begin{eqnarray}
E(SS_{E}) & = & \sum_{i=1}^{a} \sum_{j=1}^{b} \sum_{k=1}^{n} E\left(y_{ijk} - \overline{y}_{ij.}\right)^2 \nonumber \\
& = & \sum_{i=1}^{a} \sum_{j=1}^{b} \sum_{k=1}^{n} E\left(e_{ijk} - \overline{e}_{ij.}\right)^2 \nonumber \\
& = &  abn\left(\sigma^2 - \frac{\sigma^2}{n}\right) = ab(n-1)\sigma^2\nonumber
\end{eqnarray}
It follows that the posterior distribution of the variance parameter $\sigma^2$ is an inverse gamma distribution, with the $SS_{E}/2$ as the scale parameter. The variance parameter has an inverse-gamma distribution with shape parameter $(g_1+ab(n-1))/2$ and scale parameter $(g_2+SS_{E})/2$.

Then in \textbf{step 2b}, the posterior distribution of the covariance parameter $\tau_b$ is derived by determining the expected value of the $SS_{B}$, which is the sufficient statistic. It follows that,
\allowdisplaybreaks
\begin{eqnarray}
E(SS_{B}) &=& E\left(\sum_{i = 1}^{a} \sum_{j = 1}^{b} n \left(\overline{y}_{ij.} - \overline{y}_{i..}  \right)^{2}\right) \nonumber \\
& = & \sum_{i = 1}^{a} \sum_{j = 1}^{b} n E\left( (\mu + \alpha_{i} + \beta_{ij} + \overline{e}_{ij.}) - (\mu + \alpha_{i} + \overline{\beta}_{i.} + \overline{e}_{i..})\right)^2  \nonumber \\
& = & \sum_{i = 1}^{a}  \sum_{j = 1}^{b} n E\left(\beta_{ij} - \overline{\beta}_{i.}\right)^2 + nE\left(\overline{e}_{ij.} - \overline{e}_{i..}\right)^{2} \nonumber \\
& = & \sum_{i = 1}^{a}  \sum_{j = 1}^{b} n \left(\tau_b - \frac{\tau_b}{b}\right) + n\left(\frac{\sigma^2}{n} - \frac{\sigma^2}{nb}\right) \nonumber \\
& = &  a(b-1)\left(n\tau_b  + \sigma^2 \right). \label{twowaytaub}
\end{eqnarray}
The prior for the parameter $\tau_b$ is defined as
\begin{eqnarray}
p\left(\tau_b \mid \sigma^2 \right) \propto \left(\tau_b + \sigma^{2}/n\right)^{-1},
\end{eqnarray}
which allows $\tau_b$ to be negative but greater than $-\sigma^{2}/n$. \textbf{Step 3b}: The posterior distribution for $\tau_b$ is a shifted inverse-gamma distribution with shape parameter $a(b-1)/2$, scale parameter $SS_{B}/n$ and shift parameter $\sigma^{2}/n$.

\textbf{Step 2a}: the posterior distribution of the covariance parameter $\tau_a$ can be obtained in the same way, by considering the expected sum of squares of $SS_{A}$,
\allowdisplaybreaks
\begin{eqnarray}
E\left(SS_{A}\right) & = & bn \sum_{i=1}^{a} E\left((\mu + \alpha_i +\overline{\beta}_{i.} + \overline{e}_{i..}) -  (\mu + \overline{\alpha}_{.} +\overline{\beta}_{..} + \overline{e}_{...})  \right)^2 \nonumber \\
& = & bn \sum_{i=1}^{a} E\left(\alpha_i - \overline{\alpha}_{.}\right)^2 + E\left(\overline{\beta}_{i.} - \overline{\beta}_{..}\right)^2 + E\left(\overline{e}_{i..} - \overline{e}_{...}\right)^2 \nonumber \\
& = & bna\left(\left(\tau_a - \frac{\tau_a}{a}\right) + \left(\frac{\tau_b}{b}- \frac{\tau_b}{ab}\right) + \left(\frac{\sigma^2}{bn} -\frac{\sigma^2}{abn}\right)  \right)  \nonumber \\
& = & (a-1)\left(bn\tau_a + n\tau_b + \sigma^2\right). \label{twowaytaua}
\end{eqnarray}
The $SS_{A}/(bn)$ is the sufficient statistic for the $\tau_a$, then the prior for $\tau_a$ equals
\begin{eqnarray}
p\left(\tau_a \mid \tau_b, \sigma^2 \right) \propto \left(\tau_a + \left(\tau_b/b + \sigma^2/(bn)\right) \right)^{-1}.
\end{eqnarray}
\textbf{Step 3a}: it follows that the posterior distribution of $\tau_a$ is shifted inverse-gamma with shape parameter $(a-1)$, scale parameter $SS_{A}/(bn)$, and shift parameter $\tau_b/b + \sigma^2/(bn)$. The $\tau_a$ is restricted to be greater than $-(\tau_b/b + \sigma^2/(bn))$, where $\tau_b > -\sigma^2/(n)$.

\subsubsection{Multi-way Classification}
In a more general description, for a balanced design a Gibbs sampling procedure can be defined for any multi-way classification model, where different types of clustering group the continuous data. The (lower-level) variance parameter has an inverse-gamma posterior distribution, where the $SS_{E}$ is the sufficient statistic. Each covariance parameter has a shifted inverse-gamma distribution, which is constructed from the sum of squares representing the corresponding sufficient statistic. The parameter space of the variance components covers those negative values that still lead to a positive definite covariance matrix. In the Gibbs sampling algorithm, the variance components can be iteratively sampled from their posterior distributions, which leads to a very fast and efficient sampling method.

The MCMC algorithm is easily extended when including a sampling step for fixed effect parameters. Consider the BCSM in Equation \eqref{BCSM1}, and let  $\mu=\mathbf{X}_i\bm{\beta}_f$. The covariance matrix $\bm{\Sigma}$ has two parameters $\sigma^2$ and $\tau$, and the inverse of the covariance matrix is known \parencite{Searle1992}. When assuming an uniform prior, the posterior distribution for $\bm{\beta}_f$ is normal with variance and mean
\begin{eqnarray}
Var\left(\bm{\beta}_f \mid \mathbf{y},\bm{\Sigma} \right) & = & \bm{\Omega}=\left(\mathbf{X}^t \left(\mathbf{I}_a \otimes \bm{\Sigma}^{-1} \right)\mathbf{X}\right)^{-1}, \nonumber \\
E\left( \bm{\beta}_f \mid \mathbf{y},\bm{\Sigma} \right) & = & \bm{\Omega} \mathbf{X}^t\left(\mathbf{I}_a \otimes \bm{\Sigma}^{-1} \right)\mathbf{y}, \nonumber
\end{eqnarray}
respectively.

\section{Simulation Study}
The BCSM estimation method was investigated for small variance components close to the lower bound of zero, for a few clusters, few observations for each cluster, and even for negative cluster dependencies. Data was simulated under a random intercept model, with the residual variance equal to $\sigma^{2}_{e} = 5, 1, 0.5, 0.1, 0.01$, and the random intercept variance  equal to $\tau = 5, 1, 0.5, 0.1, 0.01$. The general mean was simulated from a standard normal distribution. The number of clusters was equal to $a=50,25,10,5$, and the number of observations per cluster $n=20,10,5,2$. All conditions were crossed with each other resulting in 400 simulation conditions. For each condition, 1,000 data replications were made according to the random intercept model defined in Equation \eqref{simple_conditional_model}, and they are referred to as the conditional data. The conditional data was analysed with LME4, which produced (restricted) maximum likelihood (REML) estimates for the variance components. Furthermore, an MCMC estimation method was used (JAGS), with (vague) inverse-gamma priors for the variance components (shape and scale parameter equal to .01), and a noninformative normal prior for the general mean. The median of the posterior distribution was used as a point estimator for the variance components, since these distributions were often asymmetric.

In the BCSM, the parameter $\tau$ is a covariance parameter, which can also be negative. Therefore, data was also generated with $\tau$ negative, but just above the lowerbound; $L_{b}= -\sigma^2/n+10^{-4}$, which assured that the covariance matrix was positive definite. Data was simulated under the BCSM, defined in Equation \eqref{BCSM1}, for the same conditions as described for the random intercept model. The condition $\tau=L_b$ was added for each combination of $a$, $ n$, and $\sigma^2$. This led to a total of 480 conditions. For each condition, 1,000 data sets were generated under the BCSM, and referred to as marginal data. The data was analysed with the BCSM, LME4, and JAGS. The main interest was the estimation of the (co)variance component, $\tau$.

The RMSE, bias and 95\% coverage rate (CR) was used as a criterion to evaluate the estimation results. The estimated CR represented the proportion that the true parameter value was covered by the 95\% credible interval (CI) across the 1,000 data replications and should be around the advocated 95\%. The 95\% CIs were computed using the MCMC samples. The MCMC algorithms for the BCSM and the random intercept model (JAGS) were ran for 10,000 iterations, while using 5,000 iterations as the burn-in period. The MCMC samples showed good convergence in each condition, which was inspected using the MCMC convergence tools in the \texttt{coda} \texttt{R}-package. The effective sample size was around 90\% for both MCMC methods for all model parameters.

Negative cluster dependence was simulated in the marginal data under the BCSM (Equation \ref{BCSM1} and \ref{variance_covariance_matrix}), since this could not be done with the random intercept model. The marginal data generated under the BCSM was used to evaluate the performance of the BCSM for all values of $\tau$. The conditional data generated under the random intercept model was used to evaluate the performance of LME4 and JAGS, when the true value of $\tau$ was positive.

{\color{black}
\subsubsection{Negative within-cluster correlation}
In Table \ref{tab:lower_bound_tau}, the estimation results (RMSE, bias, and CR) for the lower bound under the different models are shown. The lower bound varied across the different cluster sizes $n$ from $-$0.05 to $-$0.50, where the residual variance was equal to one. Under LME4, the $\tau$ represents the random intercept variance, and the corresponding estimates were all equal to zero for the simulated negative cluster dependencies. Therefore, the estimated average bias is equal to the lower bound, and the RMSE equal to the average bias. Under JAGS, the average bias is slightly higher, since the estimates for $\tau$ were just above zero due to the prior for $\tau$. When decreasing the number of clusters and the cluster size, the prior influence and the average bias increased. The bias hardly reduced when increasing the number of clusters. The estimated CRs under JAGS are all zero, since the 95\% CIs are restricted to positive parameter values and the true values for $\tau$ are negative. The random intercept model cannot describe negative cluster dependence, and estimation results under LME4 and JAGS show a bias equal to the true value for $\tau$.

Table \ref{tab:lower_bound_tau} shows that accurate estimates were obtained under the BCSM. The RMSE slightly increased when reducing the cluster size from 20 to 2. However, the bias was approximately zero in all conditions. Only in the extreme scenario with 10 observations across 5 clusters, the average bias was not around zero and $-$0.11. In that case, the posterior distribution for $\tau$ was skewed left, and the point estimate for $\tau$ slightly underestimated the true value. When increasing the number of clusters and/or the cluster size, the posterior was less skewed leading to more accurate point estimates and to RMSE and bias estimates of approximately zero. The CRs were around 95\% across all conditions, which shows that the posterior under BCSM accurately described the distribution for $\tau$.
}
\begin{table}[]
	\caption{RMSE, 95\% CR in brackets, and bias of estimator $\hat{\tau}$ of the lower bound $(L_{b})$ with $\sigma^2=1$.}
	\label{tab:lower_bound_tau}
	\begin{tabular}{rcclcclcclcc}
		\hline
		\multicolumn{1}{c}{$a$}   & \multicolumn{2}{c}{$n = 20$}  & & \multicolumn{2}{c}{$n = 10$} & & \multicolumn{2}{c}{$n = 5$}   & & \multicolumn{2}{c}{$n = 2$}    \\
		                                                 & \multicolumn{2}{c}{$L_b=-$0.05}  & &\multicolumn{2}{c}{$L_b=-$0.10} & & \multicolumn{2}{c}{$L_b=-$0.20} & & \multicolumn{2}{c}{$L_b=-$0.50} \\
		          & RMSE & Bias && RMSE & Bias && RMSE & Bias && RMSE & Bias      \\
                \cline{1-3} \cline{5-6} \cline{8-9} \cline{11-12}
		          & \multicolumn{10}{c}{LME4}                                   \\ \cline{2-12}
		50     &0.05       &  0.05 && 0.10       & 0.10 && 0.20       & 0.20 && 0.50  &  0.50   \\
		25     & 0.05       & 0.05 && 0.10       & 0.10 && 0.20       &0.20  && 0.50   & 0.50  \\
		10     & 0.05       &  0.05&& 0.10       & 0.10 && 0.20       & 0.20 && 0.50   & 0.50   \\
		5        & 0.05      &   0.05 && 0.10    & 0.10 && 0.20       & 0.20 && 0.50  &  0.50   \\ \\
		    & \multicolumn{10}{c}{JAGS}                                   \\ \cline{2-12}
		50   & 0.05 (.00) & 0.05 && 0.11 (.00) & 0.11 && 0.21 (.00) & 0.21 && 0.51 (.00) & 0.51\\
		25    & 0.06 (.00) & 0.06& & 0.11 (.00) & 0.11 && 0.21 (.00) & 0.21 && 0.52 (.00) & 0.52\\
		10    & 0.06 (.00) & 0.06 && 0.12 (.00) & 0.12 && 0.23 (.00) & 0.23 && 0.54 (.00) & 0.54\\
		5      & 0.07 (.00) & 0.07 && 0.14 (.00) & 0.14 && 0.25 (.00) & 0.25 && 0.57 (.00) & 0.57\\
		&              &  &            &  &            &  &            \\
		&   \multicolumn{10}{c}{BCSM}                                   \\ \cline{2-12}
		50    & 0.00 (.95) & 0.00 && 0.01 (.95) & 0.00 && 0.02 (.94) & 0.00 && 0.10 (.96) & $-$0.01\\
		25    & 0.00 (.94) &0.00  && 0.01 (.94) & 0.00 && 0.03 (.94) & 0.00 && 0.15 (.95) & $-$0.02\\
		10    & 0.00 (.95) & 0.00 && 0.02 (.94) & 0.00 && 0.05 (.93) & 0.04 && 0.26 (.94) & $-$0.00\\
		5      & 0.01 (.96) & 0.00 && 0.02 (.95) & 0.01 && 0.07 (.96) & 0.00 && 0.41 (.96) & $-$0.11 \\ \hline
	\end{tabular}
\end{table}

\subsubsection{Small variance component}
\begin{figure}[hbt!]
	\centering
	\includegraphics[width=150mm]{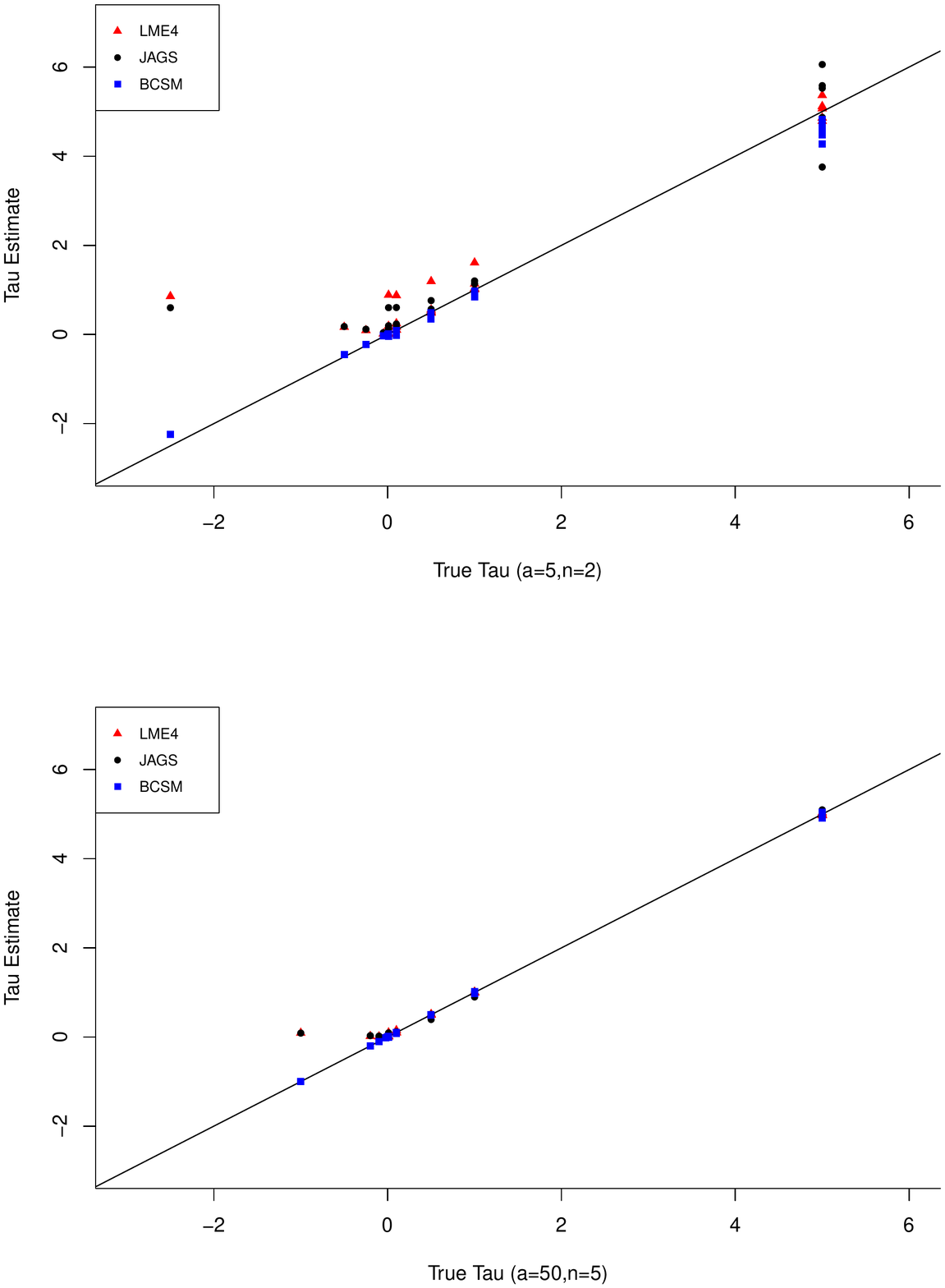}
	\caption{Averaged parameter estimates for $\tau$ across 1,000 data replications under LME4, JAGS, and the BCSM.}
	\label{fig:BCSM_introduction_estimation_true_tau}
\end{figure}

The estimation methods, referred to as LME4 (REML), JAGS, and BCSM, performed comparable in the conditions with sufficient data to estimate the parameters. When considering the $\tau$ estimates, the parameter estimates are alike, when the $\tau$ is positive (and not close to zero) and there is sufficient data, i.e. sufficient number of clusters and number of observations per cluster. In Figure \ref{fig:BCSM_introduction_estimation_true_tau}, the bottom plot shows the estimates for $\tau$ averaged across 1,000 replications under LME4, JAGS, and the BCSM in the condition with $a=50$ groups and $n=5$ observations per group with the residual variance $\sigma^2$ and $\tau$ varying across the different specified levels. For each $\tau$ value, data was generated with five different residual variances ranging from 5 to .01. In total 30 estimates for $\tau$ are plotted for each method. It can be seen that the estimates are close to the true value. In the standard situations, Figure \ref{fig:BCSM_introduction_estimation_true_tau} shows that the BCSM performs on par with the standard estimation methods.

However, when the true value is almost zero or close to zero, the BCSM outperforms the other methods. In that case, the BCSM still provides accurate estimates for $\tau$ for all values of the residual variance. The upper plot shows an extreme condition with just five groups with each two observations. It can be seen that the BCSM estimates are still close to the true value, but it tends to underestimate a true $\tau$ of five. With JAGS, the $\tau$ is overestimated in more cases when the true value is close to zero or negative. The inverse-gamma prior for $\tau$ led to an overestimation of the true value, although the hyper parameter values were .01. The REML results with LME4 also overestimated the true value, when it was negative. Furthermore, in the situation with a lack of prior information and poor data information, the REML estimates were much higher than the true values.


\subsubsection{MSE}
When considering the MSEs for the $\tau$ estimates, JAGS, LME4 and BCSM performed comparably good, when there is sufficient data information. However, for true negative values of $\tau$, the BCSM outperformed the other methods. Furthermore, the MSEs of the REML estimates are much higher, when the residual variance is equal to five. In the small sample conditions, the BCSM outperformed both other methods. The MSEs under JAGS and LME4 are higher when the true value is close to zero or negative. When the residual variance is five, the MSEs under LME4 are also much higher than for the other methods. For all the considered conditions, the MSEs under the BCSM are close to zero.

\subsubsection{Coverage}
Finally, the 95\% CRs were computed under JAGS and the BCSM method. CIs for the variance components were not computed under maximum likelihood estimation (LME4), since this led to numerical problems and invalid CIs (e.g. using bootstrap function in LME4 to compute CIs). The estimated CRs for JAGS were zero, when the true value was negative. The inverse-gamma prior for $\tau$ restricted the posterior distribution of $\tau$ to only cover positive values, which led to incorrect CRs. Under JAGS, the CRs were too large and close to one, when the true value of $\tau$ was positive but close to zero, and when there was not much data. In the situation without much data information, the posterior was more stretched by the prior which led to wider CIs than expected under the data replications.

The data were generated for fixed values for $\tau$, which means that the prior variance was not included in the data replications under the BCSM and random intercept model. This led to an overestimation of the 95\% CRs in the extreme data conditions, when the prior variance influenced the width of the CIs. For instance, for JAGS, with $\sigma^2=1$ and $\tau=.01$ the estimated 95\% CRs were around .98$-$1, but mostly one for all considered samples sizes. In general, accurate 95\% CRs were computed for the BCSM. Only in the extreme conditions, for instance when the number of clusters was five, the estimated CRs were close to one for the BCSM. For JAGS, the 95\% CRs were more often overestimated also for conditions with more than five groups. The inverse-gamma prior for $\tau$ in JAGS influenced more the 95\% credible regions and gave more weight to higher values than the shifted-inverse gamma prior in the BCSM. This is a typical issue for inverse-gamma priors for variance components in hierarchical models \parencite{Gelman2006a}. In the BCSM, the shifted-inverse gamma prior performed better simply by extending the parameter space to include negative values.

\section{Personalized Treatment in E-mail Counselling}

The effectiveness of BCSM is demonstrated for a real-data example. \textcite{Lamers2015} examined whether a combination of a self-help intervention with narrative therapy is effective in alleviating symptoms of depression and anxiety. The treatment consisted out of two conditions: the \textit{auto-biographic} and the \textit{expressive} writing condition (AW and EW, respectively). The AW condition was a life-review self-help intervention that consisted of homework assignments, divided over modules that had to be completed over the course of ten weeks. Clients communicated about their progress with trained counsellors through a weekly e-mail interaction. The EW intervention was based on the method of expressive writing. The method consisted of daily writing about emotional experiences, for $15-30$ minutes on $3-4$ consecutive days during one week. \textcite{Lamers2015} used a repeated measures ANOVA and found that depressive symptoms indeed declined, but did not find a difference between the AW and EW condition (in comparison with a waiting list control group). \textcite{Smink2019_Frontiers} adopted a multilevel approach with client as a random effect, and also did not find a significant difference between treatments.

The BCSM was used to identify individual variability in treatment effects and to identify those who benefitted from the treatment, since a significant main treatment effect could not be found. Furthermore, the object was to investigate the effect of the counsellor and how they contributed to the treatment of the clients. Several clients in different treatment arms were treated by the same counsellor, and negative clustering effects were expected since individualized treatments were given by each counselor. Let $i$ denote the index for the counsellor and $j$ the client. Each client was measured at a pre- and a post-intervention occasion, which resulted in a $y_{ij1}$ and $y_{ij2}$ score, respectively. Scores of clients who were treated by the same counsellor (i.e. counsellor $i$) were assumed to be clustered, and we also assumed that scores coming from the same client (i.e. client $j$) were clustered. Let factor variable $\alpha_i$ represent the counselling effect, and nested factor variable $\beta_{(i)j}$ the client effect. This leads to a two-way nested factor model for the pre- and post-intervention scores presented in Equation \eqref{twowayconditional}. In the corresponding BCSM, the covariance structure implied by the two factor variables was directly modelled, allowing for the occurrence of potentially negative cluster correlations.

\subsection{Measuring client, counsellor and individual treatment effects}
We first fitted a linear regression model, denoted as LM M0, which assumed independently distributed errors. Then, two BCSMs were considered, to which we refer as M1 and M2, which had the same mean term as the LM M0. For all three models, the intercept $\beta_0$ represented the average score at the pre-intervention for clients in the AW condition. The treatment variable was dummy-coded (with a \textit{one} for clients in condition EW, and a \textit{zero} for those in condition AW). The main effect of treatment, represented by $\beta_1$, was included to correct for any pre-intervention differences between the two treatment groups. The $\beta_2$ represented the average contribution of the post-intervention in comparison to the pre-intervention score, where indicator variable $Post_{ij}$ was also dummy-coded (with a \textit{one} for the post-intervention scores, and \textit{zero} for the pre-intervention scores). An interaction variable $Z$ with effect $\beta_3$ was dummy coded, where a \textit{one} represented the interaction between the post-intervention measurement of clients in condition EW.

BCSM M1 and M2 assumed dependence among scores from clients assigned to the same counsellor, and M2 also assumed a dependence among pre- post-intervention scores from the same client. To better understand the factor structure represented in the covariance structure of BCSM M2, consider the (conditional) MLM with random effects for the counsellor and the client;
\begin{eqnarray}
y_{ijl} & = & \beta_0 + \beta_1Treatment_{ijl} + \beta_2{Post}_{ijl} + \beta_3Z_{ijl} + \alpha_i + \beta_{(i)j} + e_{ijl}, \nonumber \\
\alpha_i & \sim & N(0,\tau_a) \textrm{ (Counsellor)}   \nonumber \\
\beta_{(i)j} & \sim & N(0,\tau_b) \textrm{ (Client)}  \nonumber \\
e_{ijl} & \sim & N(0,\sigma^2), \nonumber
\end{eqnarray}
where $l=1,2$ indicates a pre-intervention or post-intervention observation, respectively. The MLM cannot detect negative clustering effects, and it needs 90 random client parameters and five random counsellor parameters to model the dependence structure. This makes it unsuitable for the small data set. Therefore, the dependence structure is directly modeled, which leads to the following BCSM:
\begin{eqnarray}
\mathbf{y}_{i} & = & \beta_0 + \beta_1\textbf{Treatment}_i + \beta_2\textbf{Post}_{i} + \beta_3\textbf{Z}_i + \mathbf{E}_{i}, \label{BCSMRD}
\end{eqnarray}
where the $\mathbf{E}_{i}$ is (multivariate) normally distributed. The three models --LM M0, BCSM M1, and BCSM M2-- can be represented by the model in Equation \eqref{BCSMRD}, but each model has its specific (structured) covariance matrix $\bm{\Sigma}$. For model LM M0, the covariance matrix $\bm{\Sigma}=\sigma^2\mathbf{I}_{n}$ represents independently distributed errors. For BCSM M1, a one-way clustering is assumed represented by the covariance matrix $\bm{\Sigma}=\sigma^2\mathbf{I}_n+\tau_a\mathbf{J}_n$. For BCSM M2, the covariance matrix $\bm{\Sigma}$ is given in Equation \eqref{twowaySigma}. The covariance matrix of M1 represents a one-way clustering with $\tau_a$ the covariance of scores of those treated by the same counsellor. The covariance matrix of BCSM M2 also includes a component $\tau_b$, which represents the covariance of scores of the same client.

For the BCSM models M1 and M2, an MCMC algorithm with 20,000 iterations (with a burn-in of 1,000 iterations) was used to compute the parameter estimates. The parameter estimates of BCSM M1 and M2 are given in Table \ref{tab:Lamers_et_al_2015_data1}. The BCSM M2, with a two-nested dependence structure, contained four regression parameters and three (co)variance parameters. This makes the BCSM particularly useful for small data sets. A trimmed mean estimator was used for the covariance components, where 10\% of the outlying values were ignored to obtain more robust posterior mean estimates. The posterior standard deviations were estimated using all sampled values. The parameter estimates of model M0 were obtained using the \texttt{lm}-function in \texttt{R}. {\color{black} In the Supplementary Materials Section D, next to the mean and standard deviation estimates in Table \ref{tab:Lamers_et_al_2015_data1}, the 95\% highest posterior density intervals have also been added for the parameters of the BCSMs.}

\begin{table}[]
	\caption{The e-mail-counselling study from \textcite{Lamers2015}: A BCSM analysis of the pre- and post-intervention data.}
	\label{tab:Lamers_et_al_2015_data1}
	\begin{tabular}{lllclclrlr}
		\hline
		&                      &                      & LM ($\hat{M}$, $S.E.$)             &                      & \multicolumn{5}{c}{BCSM ($\hat{M}$, $SD$)}                                                                                         \\ \cline{4-4} \cline{6-10}
		\multicolumn{1}{c}{} & \multicolumn{1}{c}{} & \multicolumn{1}{c}{} & M0                                 & \multicolumn{1}{c}{} & M1                                 & \multicolumn{1}{c}{} & \multicolumn{1}{c}{M2} & \multicolumn{1}{c}{} & \multicolumn{1}{c}{M3} \\ \cline{4-4} \cline{6-6} \cline{8-8} \cline{10-10}
		\multicolumn{2}{l}{\textit{Fixed effect}}   &                      &                                    &                      &                                    &                      & \multicolumn{1}{c}{}   &                      & \multicolumn{1}{c}{}   \\ \cline{1-2}
		Intercept            & $\beta_{0}$          &                      & \multicolumn{1}{r}{21.78 (0.91)}   &                      & \multicolumn{1}{r}{21.72 (0.84)}   &                      & 21.67 (0.79)           &                      & 21.68 (0.80)           \\
		Treatment            & $\beta_{1}$          &                      & \multicolumn{1}{r}{$-$0.29 (1.29)} &                      & \multicolumn{1}{r}{$-$0.16 (1.29)} &                      & $-$0.08 (1.29)         &                      & $-$0.09 (1.31)         \\
		Post                 & $\beta_{2}$          &                      & \multicolumn{1}{r}{$-$4.04 (1.28)} &                      & \multicolumn{1}{r}{$-$4.06 (1.29)} &                      & $-$4.03 (0.99)         &                      & $-$4.05 (1.01)         \\
		Interaction          & $\beta_{3}$          &                      & \multicolumn{1}{r}{$-$1.36 (1.81)} &                      & \multicolumn{1}{r}{$-$1.33 (1.82)} &                      & $-$1.37 (1.42)         &                      & $-$1.35 (1.43)         \\
		&                      &                      &                                    &                      &                                    &                      & \multicolumn{1}{c}{}   &                      & \multicolumn{1}{c}{}   \\
		\multicolumn{2}{l}{\textit{Random effects}} &                      &                                    &                      &                                    &                      & \multicolumn{1}{c}{}   &                      & \multicolumn{1}{c}{}   \\ \cline{1-2}
		Residual             & $\sigma^{2}$         &                      & 37.04                              &                      & \multicolumn{1}{r}{37.79 (4.03)}   &                      & 21.73 (3.32)           &                      & 22.08 (4.89)           \\
		Counsellor           & $\tau_{a}$           &                      &                                    &                      & \multicolumn{1}{r}{$-$0.68 (0.44)} &                      & $-$1.12 (0.49)         &                      & $-$1.07 (0.51)         \\
		Client               & $\tau_{b}$           &                      &                                    &                      &                                    &                      & 15.83 (4.52)           &                      & 15.54 (4.74)           \\
		Interaction          & $\tau_{c}$           &                      &                                    &                      &                                    &                      & \multicolumn{1}{c}{}   &                      & 6.28 (7.79)            \\ \hline
	\end{tabular}
\end{table}

The parameter estimates of the regression effects did not differ much for the different models. The adjusted $R^2$ was around .91 under model LM M0. It can be seen that on average on the post-intervention clients scored four points lower than on the pretest, showing that depressive symptoms indeed declined. There were no significant differences between the two treatment groups on the pre-intervention. The interaction effect $\beta_3$ was around $-$1.36, showing that those in the EW condition scored on average lower than those in the AW condition at the post-intervention. However, the posterior probability of a negative interaction effect $P(\beta_3<0 \mid \mathbf{y})$ was around 84\% under M2. There was no convincing data evidence that on average the EW treatment outperformed the AW treatment.

When interpreting the estimated covariance components under M1 and M2, it can be seen that the estimated covariance among scores of clients assigned to the same counsellor was negative under the BCSM models, and around $\tau_a=-.68$ under M1. Thus, scores from clients treated by the same counsellor correlated negatively. This led to an increase of the residual variance estimate for M1 in comparison to the estimated residual variance of model M0. The residual variance was underestimated under M0, since the residuals were not independently distributed but correlated negatively. The standard deviation of the intercept was around 8\% smaller under M1 in comparison to M0. The negative correlation among client scores affected the estimated standard deviation of the intercept, where the standard deviations of the other regression components under M1 were almost equal to the corresponding standard errors under M0. The dependence structure implied by the clustering of clients by counsellors cannot be represented by a counsellor random effect, since the estimated cluster correlation was negative. This makes the BCSM particularly useful to model negative cluster correlation.

When accounting for the dependence among client's pre- and post-intervention scores, the estimated covariance of $\tau_a$ was more negative under M2 than under M1 and around $-$1.12. This led to a further reduction of the standard deviation of the intercept to .79. This negative covariance of $\tau_a$ led to an increase of the residual variance under M2. However, the estimated positive covariance of $\tau_b=15.83$ led to a decrease of the residual variance to 21.73. The estimated standard deviation of the Post effect, $\beta_2$ decreased to .99 due to accounting for the correlation between client's scores. The standard deviation of the interaction effect also seriously decreased from 1.82 to 1.42. Note that the dependence structure did not influence the standard deviation of the pre- post-intervention difference between treatment groups (i.e. standard deviation of $\beta_1$).

The negative correlation among client scores from the same counsellor indicated that there was individual variability in treatment effects across the clients of the same counsellor. In the same condition and for the same counsellor, some clients benefited from the treatment, where others did not and even showed an increase in score. This phenomenon of individual treatment effects was identified by the negative cluster correlation, which was also significant when considering the 95\% HPD interval under M1 and M2. The negative cluster correlation of $-$1.12 illustrated that there was more heterogeneity in test scores than explained by the reduction in scores at the post-intervention and the (non-significant) mean difference between the two conditions.

\subsection{Post-intervention individual treatment effects}
To investigate the individual treatment effect further, the model BCSM M3 was defined with a random interaction effect. This represented random variability in the treatment condition EW at the post-intervention across clients, while also accounting for the clustering by clients and counsellors. The covariance structure of M3 for the client scores of counsellor $i$ is given by
\begin{eqnarray}
\bm{\Sigma}_i & = & \mathbf{I}_{nb}\sigma^2 + \underbrace{diag(\mathbf{Z}_{i})\tau_c}_{\text{Interaction}} + \underbrace{\mathbf{J}_{nb}\tau_a}_{\text{Counsellor}} + \underbrace{\left(\mathbf{I}_b\otimes\mathbf{J}_n\right)\tau_b}_{\text{Client}}, \label{BCSMM3}
\end{eqnarray}
and the covariance matrix is counsellor specific due to the $\mathbf{Z}_{i}$. However, this random interaction-effect cannot be estimated, since each client only had one observation at the post-intervention. The interaction variable $\mathbf{Z}_{i}$ is a diagonal matrix in the covariance matrix with $\tau_c$ a residual variance parameter. Thus, the random interaction effect implies an interaction-specific residual variance in the covariance matrix. The dependence structure in Equation \eqref{BCSMM3} represents heteroscedastic residual variances, with $\sigma^2$ the common residual variance and $\sigma^2_1 =\sigma^2+\tau_c$ the contribution of the random interaction variance to the common residual variance. Note that the dependence structure is extended with just one additional variance parameter representing the random interaction variance for clients in the EW condition. There is data evidence in favor of individual treatment effects of clients in the EW condition, when the residual variance $\sigma^2_1$ is greater than $\sigma^2$. In Section C of the Supplementary Materials, the posterior distribution of $\sigma^2$ and $\tau_c$ is given, and the adjustment of the shift parameters in the posterior distributions of the other covariance parameters.

The estimates of BCSM M3 are given in Table \ref{tab:Lamers_et_al_2015_data1}. It can be seen that the fixed regression effects did not change when including the random interaction effect. The estimated residual variance was slightly higher. The standard deviation increased, since less observations were used to estimate the common residual variance. The estimated cluster dependence of clients and of counsellors were also around the estimated values of BCSM M2. The estimated random interaction variance was around $6.28$, which showed that there was more residual variance in the post-intervention scores in the EW condition. In the Supplementary Materials Section C it is shown that the $\tau_c \ge -\sigma^2$, and $\tau_c$ is allowed to be negative. The BCSM simply makes it possible to evaluate the data support in favor of individual variation, since the interaction variance is allowed to be negative. In this case, the interaction variance was estimated to be positive with 80\% posterior probability.

The effect of the EW-treatment varied across individuals, where some benefitted more from the treatment than others. The relatively large individual variation showed that for some clients the EW-treatment was very effective but not for others. A main difference between treatments was not found partly due to this individual variation. The posterior standard deviation of the interaction variance was high and around 7.79, and around 20\% of the posterior distribution of the $\tau_c$ supported negative variance values. In that case, the random interaction effect lead to a common reduction in the residual variance in the EW condition, which provide more support for a main treatment effect and less support for individual variation in the treatment effect. However, an effect of a negative variance on the standard deviation of the mean interaction effect would be very small, since this can only be accomplished through the covariance matrix of the fixed effect, where it would be absorbed by other more influential factors. Nevertheless, it can be argued that $80\%$ posterior probability is sufficient to conclude that there is individual variation in the EW-treatment effect.

\subsection{Visualizing individual treatment effects}
\begin{figure}[hbt!]
	\centering
	\includegraphics[width=150mm]{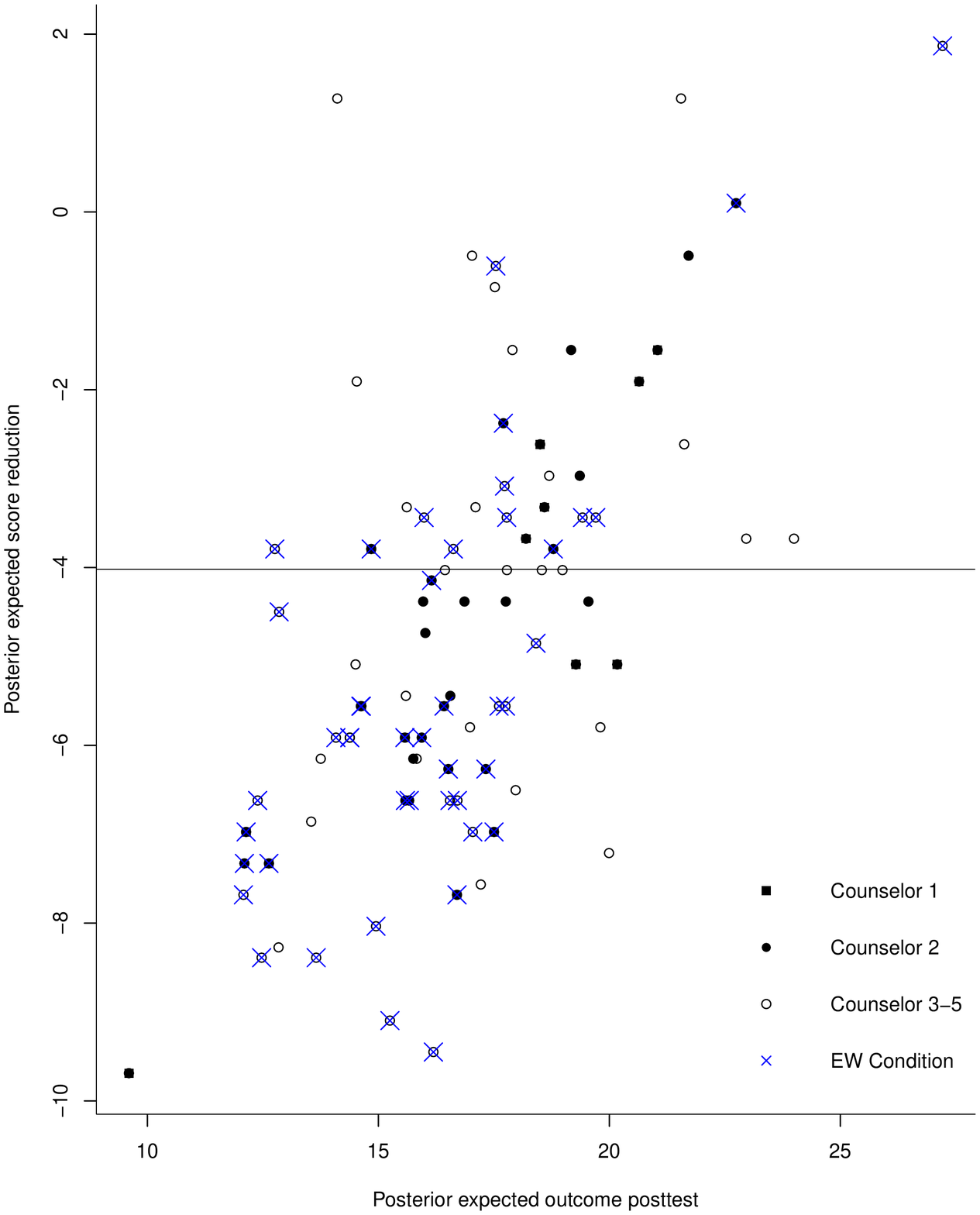}
	\caption{The email-counselling study from \textcite{Lamers2015}. Posterior expected post-intervention scores against the expected reduction in scores of clients across counsellors.}
	\label{fig:BCSM_introduction_negative_clustering_Lamers_2015}
\end{figure}

The individual variation in treatment effect is further illustrated. In Figure \ref{fig:BCSM_introduction_negative_clustering_Lamers_2015}, the posterior expected post-intervention scores are plotted against the expected reduction in scores for clients treated by different counsellors. It is shown that for counsellor 1 (filled squared box) and for counsellor 2 (filled circles), some clients show a reduction below the average of -4, where other clients treated by the same counsellor show an above-average reduction in scores. Clients treated by the same counsellor show a large deviation in reduced scores. This heterogeneity in reduced scores for clients of the same counsellor is manifested by a negative cluster correlation. This means that the level of score reductions varies across clients of the same counsellor. Therefore, it is not possible to identify a common counsellor effect, since this would imply less heterogeneity in reduced scores and a positive correlation. In fact, the effect of the counsellor varies across clients, where some clients benefitted more from the counsellor than others. This can be identified as the detection of an individualized counsellor effect. The crossed marks in Figure \ref{fig:BCSM_introduction_negative_clustering_Lamers_2015} represent reduced scores of the ES treatment. Although, reduced scores from clients in the ES condition of counsellor one are all below the average, some clients of counsellor two scored above average in this condition. The ES treatment is likely to be more effective for counsellor one than for counsellor two. Therefore, the individualized treatment effects of the counsellors may also include heterogeneity in the AW and EW treatments.


\begin{figure}[hbt!]
	\centering
	\includegraphics[width=150mm]{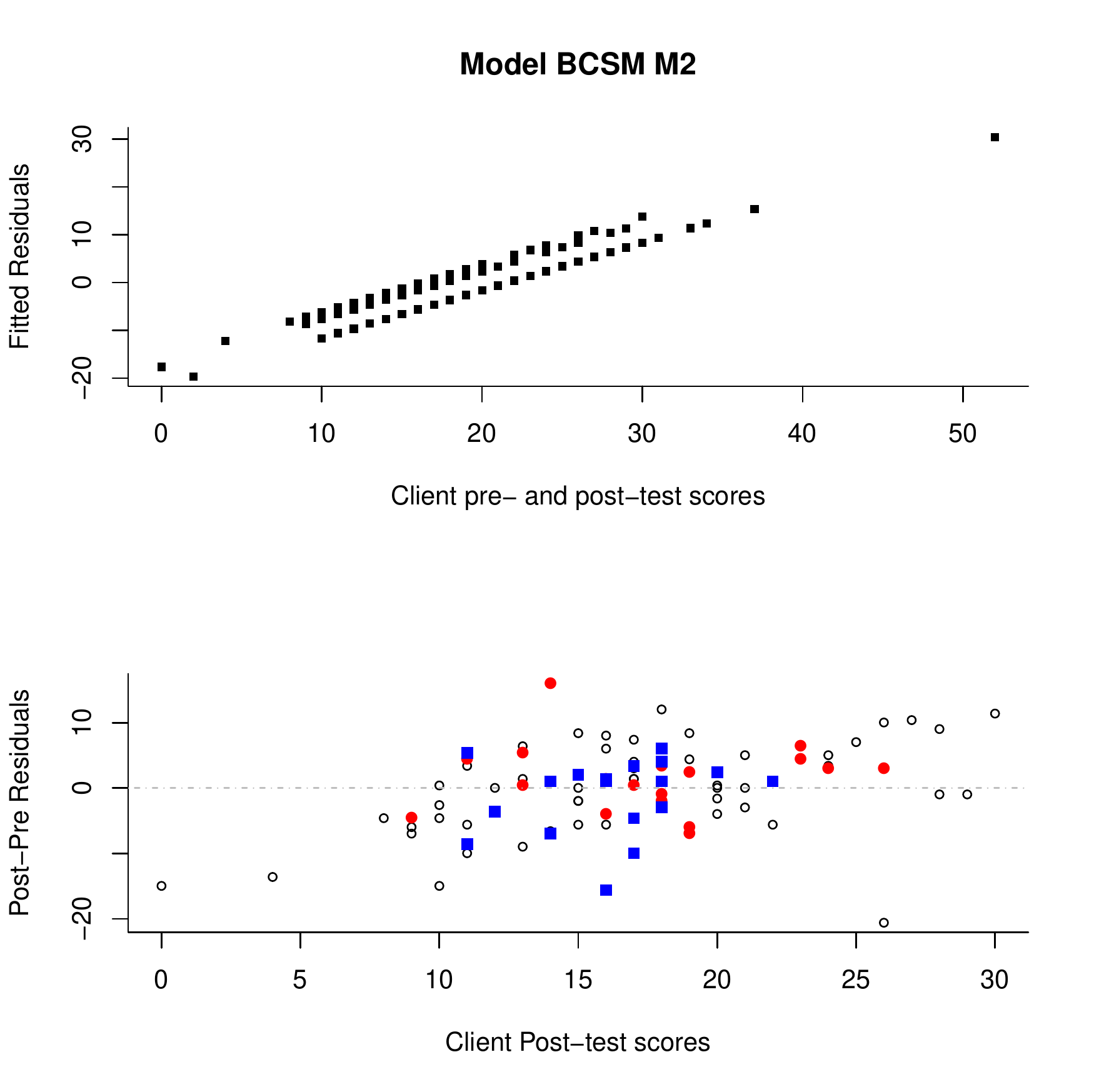}
	\caption{The fitted residuals under the BCSM M2, and the post-intervention scores against client's post- minus pre-intervention residuals for different counsellors.}
	\label{fig:BCSM_introduction_residuals_Lamers_2015}
\end{figure}

In Figure \ref{fig:BCSM_introduction_residuals_Lamers_2015}  (upper plot), the pre- and post-intervention scores are plotted against the fitted residuals under BCSM M2. It can be seen that the residuals are directly defined in relation to the outcome variable, and differences between residuals are caused by the effects of categorical predictor variables. This illustrates that the BCSM is a parsimonious model. Despite the complex two-way nested clustering structure, the fitted residuals can be directly explained by the differences caused by the categorical predictor variables. Under a latent variable model, the fitted residuals would have been scaled in relation to the estimated latent variables. The lower plot shows the post-intervention scores against the difference between the post- and pre-intervention residuals. The filled circles are those related to counsellor one, and the filled squares to those of counsellor two. It can be seen that for both counsellors, some clients showed a large decrease in residual value, where others did not. This heterogeneity across clients treated by the same counsellor in residual reduction from the pre-intervention to the post-intervention shows again that some clients benefited from the treatment, where others did not.


This analysis of the treatment effects was not possible with an MLM, since the factor variable counsellor implied a negative cluster correlation. This led to singular model, when using the LME4 package in R, and the random effects structure was considered too complex to be supported by the data. However, by ignoring the negative cluster correlation, relevant information was ignored. Counsellors provided individual instructions to their clients, which led to a decrease in scores for some clients but not for others. The differential treatment by counsellors was identified by the negative cluster correlation. The Bayesian estimation procedure for the BCSM did not have any issues in estimating the model parameters despite any negative clustering effects and the small sample size.

\section{Discussion}
We introduced the novel statistical modeling framework Bayesian Covariance Structure Modelling and emphasized the understanding of BCSM, rather than discussing the underlying mathematical rigour. We designed a simulation study and analysed real data to demonstrate that BCSM can 1) assess (very) small variance components (i.e. near the lower-bound of zero), 2) assess negative variance components, 3) assess complex dependence structures given small data sets, and 4) assess individualized effects (by modelling negative associations between clustered observations). We discuss our findings, reflect on the limitations of our study, and suggest further BCSM research.

MLM software programs can produce negative variance estimates, which in general is considered to be an objectionable characteristic of the estimation  methods, and limits the usefulness of variance component techniques \parencite{Thompson1962}. For instance, the online SAS documentation (section Negative Variance Component Estimates) reports that it is common practice to treat negative variance components as if they are zero (assuming the model is appropriate for the data, see \url{https://support.sas.com/en/documentation.html}). It is argued that a larger sample size might be needed, outliers cause violations of model assumptions, or the variability is too large. However, it is also stated that negative variance estimates can indicate that clustered observations are negatively correlated. The BCSM gives support to modeling negatively correlated observations while using a very parsimonious modeling approach to make it suitable for very small data sets. From a statistical point of view, the BCSM is the natural extension of the MLM approach.

{\color{black}
The MCMC algorithms for the BCSM were implemented in R, which were used for the simulation and real-data study. It was not possible to use general-purpose (Bayesian) software, such as Stan or Jags, to fit the BCSM. The shifted-inverse gamma posterior distributions for the covariance parameters are not standard. Furthermore, the MCMC algorithms include parameter restrictions, where covariance parameters are restricted by the values of other covariance parameters such that the restrictions change across MCMC iterates. More research is needed to make the BCSM software easily accessible.
}

Many statistical models rely on multiple observations for proper model behaviour. Statistical modelling runs into problems when there are only a few observations (i.e. when data is sparse), yet, small samples are by no means a rare occurrence in many scientific disciplines. After all, a (relatively) small(er) data set does not imply a lesser degree of importance, as there are a variety of reasons why data sets could be small. Correct statistical modelling is arguably even more important when, for example, the population of the target group is extremely sparse (e.g., babies with a life-threatening orphan disease), difficult to access (e.g., toddlers with autism from refugees), or very costly (e.g., heart-lung transplants in infants). Small data sets are especially challenging for mixed effects models, as the sample size restrictions apply to each (modelled) hierarchical level in the data. Limited sample sizes greatly constrain meaningful statistical inference, as the sample determines the sufficient number of clusters (usually too few), and the size of the clusters themselves (usually too small). To overcome these issues, researchers often simplify their hypotheses and corresponding statistical models. Instead of doing that, our simulation study showed that BCSM can deal with few clusters with a small number of observations.

The reason why small and even negative variance components are easily estimated under BCSM, is because the so-called boundary effects can be weakened by extending the parameter space to include negative values. Usually, zero is the lower-bound of variance components because --in the standard multilevel modelling approach-- a random effect is used to model dependences among treated individuals. However, the random effect variance is restricted to be positive and, as a result, always implies a positive association among individuals. Negative associations among measurements caused by the cluster (such as the counsellor, or the teacher), which increases the heterogeneity among treated individuals, would require the modelling of a negative random effect variance. Under BCSM, it is straightforward to assess these effects. The covariance structure of the BCSM can represent a random effect structure, but the random effects themselves do not have to be estimated. Many individual change phenomena can be represented through a multilevel model, but these methods typically require large samples and cannot always properly model heterogeneity within clusters. An important advantage of BCSM is that the covariance structure can represent a dependence structure implied by random effects, but the effects themselves do not have to be estimated. The number of BCSM parameters is drastically lower than for the standard MLM approaches, while the interpretation does not change. Thus, BCSM allows for modelling complex theories with limited data.

\subsection{Limitations}

The main limitation is that data was assumed for a balanced design with a one-way or two-way random effects structure. This textbook-case is --indeed-- simple, but also illustrative. We choose these (balanced) dependence structures to align with our ambition to also gently introduce the BCSM. The balanced design greatly simplifies the mathematical structure underlying our analyses. Ultimately, it is also our goal to include unbalanced designs, but to keep the scope of our current article manageable, we focused on balanced designs. Of course, because unbalanced designs are so ubiquitous in practice, the BCSM is going to be extended to unbalanced designs. Furthermore, the statistical results obtained for balanced designs will be the building blocks for unbalanced designs. Meanwhile, BCSMs have been defined for much more complex dependence structures \parencite[as can be seen in][]{Mulder2019, Fox2017, Klotzke2019a, Klotzke2019b}.

Another limitation is that we relied on the default settings of the LME4 and JAGS's estimation method. We could have also adjusted and tweaked the estimation methods for optimal performance. All things considered, we justified our choice based on the relative simplicity of the one-way random effects model. The methods should be able to perform equally well (without any adjustments) for these kind of models.

A final limitation lies within the computational efforts that are needed to estimate BCSM parameters. The Gibbs sampling procedure simply requires more computation time than standard maximum likelihood methods. Ultimately, we feel that the fact that the BCSM can estimate negative cluster correlations for relative small samples far outweighs the computational cost. Also, while this generally true for all analysis of data: data collection (usually) takes way more time, outweighing the computational time (usually) by a large margin.

\subsection{Future research}

The future of research into BCSM appears to be very relevant for various long-standing statistical modeling problems. One of the foremost, is model selection. As we have shown, under BCSM, zero is `just' another value in the parameter space of the (co)variance parameter instead of (an absolute) lower bound. In the (standard) MLM, inferences about random effect variance parameters are problematic. For instance, a random-effect variance of zero, or a negative variance estimate, can be of specific interest, but is now non-testable as both these values lie outside the boundary of the parameter space. Central to psychological research is that theories or hypotheses are often expressed in the form of several competing models \parencite{Klugkist2010, Wagenmakers2004}. It is also often complicated to compare models that have small variance components, as these variances lie near the lower bound, and testing near (or on) the lower-bound is known to be problematic. With the BCSM, these so-called boundary effects can be avoided, or at least weakened, by extending the parameter space to include negative values, allowing not only for a more direct, but also testable model comparison. In Bayesian hypothesis testing, hypotheses are restricted to the parameter space of the prior(s). Thus, a major improvement of the BCSM is the simple solution to have a prior distribution which gives positive support to negative and positive intra-cluster correlations to make an objective decision about the nature of the clustering.

Another interesting line of future research into BCSM is an extension to make statistical inferences from (very) small data samples: BCSM has minimal sample size requirements, since it only requires two observations to estimate the intra-cluster correlation, which is --indeed-- the bare minimum of observations required to compute a variance component. Data sets in the social and medical sciences often remind us that not all data is `Big Data': small samples are by no means a rare occurrence. A small data set does not imply a lesser degree of importance, as there are a variety of reasons why data sets could be small. Correct statistical modelling is perhaps even more important when, for example, the population of the target group is extremely sparse. Even for small data sets, researchers in the social and medical sciences often have comprehensive theories available, which lead into the direction of testing many parameters with multiple and complex dependencies. Fortunately, the complexity of the BCSM is easily controlled, since each random effect structure is modelled in a separate layer of an additive covariance structure. Doing so is much more difficult in the MLM approach, where each random effect introduces many model parameters and the exact number of parameters depends on the fit of the model.

The final suggestion for future research concerns the estimation of individual treatment effects. It is shown that the BCSM can detect individualized treatments through negative intra-individual correlations, a next step is the estimation of the effects. Estimated BCSM residuals contain the individual-specific regression (random effect) parameters and a post-hoc estimation method is needed to estimate those random effects. For positively correlated clustered observations, these estimated effects should resemble the random effect estimates under the MLM. For negatively correlated observations, a different method is needed to estimate the individual-specific contribution.

\subsection{Conclusion}

We hope that we have been able to show how our BCSM approach contribute to standard multilevel modelling approaches and can be applied to evaluate individualized interventions in psychology. Even though --as \textcite{Pryseley2011} pointed out-- negative variance components received attention for more then half a century \parencite[starting with][]{Chernoff1954, Nelder1954}, BCSM is a new way to model directly dependences between measurements and individuals.

We strongly feel that BCSM affords the possibility of estimating rich and realistic models for psychotherapy data. Given the relative importance of this question in the psychology science, we hope that the BCSM accelerates relates research into the question of how individuals change. We hope that the BCSM that we suggested serve as a starting point for empirical analyses of individual change process research, ultimately to the benefit of not only psychological science, but especially to those that rely on the benefits of (psycho)therapy.

\section{A. Covariance structure with correlated level-1 errors}
The covariance structure of the one-way random effects model is a covariance matrix with a common covariance $\tau$ plus the residual variance $\sigma^2$ on the diagonal. This additive sum of two components is based on the common covariance $\tau$ in Equation (2) and the common variance $\sigma^2$. This is known as a compound symmetry structure, which is defined in Equation (6). When the level-1 errors are correlated, the covariance structure is represented by the common covariance of clustered observations and the covariance of level-1 residuals. Assume the level-1 residuals in cluster $i$, $\mathbf{A}\mathbf{E}_{i}$, are multivariate normally distributed with covariance matrix $\bm{\Sigma}=\sigma^2\mathbf{A}\mathbf{A}^t$. Then, the outcome $\mathbf{y}_{i}$ of cluster $i$ is expressed as the sum of the general mean, $\mu$, random effect $\alpha_{i}$,  and residual errors $\mathbf{A}\mathbf{E}_{i}$,
\begin{eqnarray}\label{simple_conditional_modelA}
\mathbf{y}_{i} &=& \mu + \mathbf{1}_n\alpha_{i} + \mathbf{A}\mathbf{E}_{i}, \\
\alpha_{i} &\sim& N(0,\tau), \nonumber \\
\mathbf{E}_{i} &\sim& N(0,\sigma^2\mathbf{I}_n). \nonumber
\end{eqnarray}
The covariance matrix of the clustered observations is represented by
\begin{eqnarray}\label{covariance_i_lA}
Var\left(\mathbf{y}_{i} \right) &=& Var\left(E\left(\mathbf{y}_{i}\mid \alpha_i\right)\right) + E\left(Var\left(\mathbf{y}_{i} \mid \alpha_i \right)\right), \nonumber \\
& = & Cov\left(\mu + \mathbf{1}_n\alpha_{i},\mu + \mathbf{1}_n\alpha_{i}\right) + Var\left(\mathbf{A}\mathbf{E}_{i}\right), \nonumber \\
& = &  Cov\left(\mathbf{1}_n\alpha_{i}, \mathbf{1}_n\alpha_{i}\right)  + \mathbf{A}\sigma^2\mathbf{I}_n\mathbf{A}^t, \nonumber \\
& = & Var\left(\alpha_i \right)\mathbf{J}_{n} + \sigma^2\mathbf{A}\mathbf{A}^t = \tau\mathbf{J}_n + \sigma^2\mathbf{A}\mathbf{A}^t.
\end{eqnarray}
The covariance structure remains to be in additive form, which supports the modeling of the common covariance with a BCSM. \textcite{Klotzke2019a} discussed a BCSM for heteroscedastic level-1 residuals.

\section{B. Parameter restrictions two-way nested ANOVA}
The covariance matrix of the two-way nested model represented in Equation (8) needs to be positive definite, and this restriction leads to a lower bound for the covariance parameters $\tau_a$ and $\tau_b$. The restrictions can be obtained from the expression for the determinant of the covariance matrix. Consider the (conditional) two-way random effects model:
\begin{eqnarray}
\mathbf{y}_i & = & \mu + \left(\mathbf{I}_{b}\otimes\mathbf{1}_n\right)\bm{\beta}_{(i)} + \mathbf{1}_{nb}\alpha_i + \mathbf{e}_i.  \label{design2way1}
\end{eqnarray}
Without conditioning on the random effect parameters $\alpha_i$ and $\beta_{(i)}$, the covariance matrix of $\mathbf{y}_i$ is represented by,
\begin{eqnarray}\label{twowaySigmaB}
\bm{\Sigma} & = & \left(\mathbf{I}_{b}\otimes\mathbf{1}_n\right) Var\left(\beta_{(i)}\right) \left(\mathbf{I}_{b}\otimes\mathbf{1}_n\right)^t + \mathbf{1}_{nb}Var\left(\alpha_i\right)\mathbf{1}^t_{nb} + Var\left(\mathbf{e}_i\right) \nonumber \\
& = & \left(\mathbf{I}_{b}\otimes\mathbf{1}_n\right)\left(\mathbf{I}_{b}\otimes\mathbf{1}_n\right)^t\tau_b + \mathbf{1}_{nb}\mathbf{1}^t_{nb}\tau_a + \mathbf{I}_{nb}\sigma^2 \nonumber \\
& = &  \left(\mathbf{I}_b\otimes\mathbf{J}_n\right)\tau_b + \mathbf{J}_{nb}\tau_a + \mathbf{I}_{nb}\sigma^2\label{design2way2}
\end{eqnarray}
\textcite{LaMotte1972} gives an expression for the determinant of a nested covariance matrix. Given the design matrices of the two-way random effects model in Equation (\ref{design2way1}), the determinant is given by
\begin{eqnarray}
\left|\bm{\Sigma} \right| & = & \left(nb\tau_a + n\tau_b + \sigma^2 \right)\left(n\tau_b+\sigma^2 \right)^{b-1}\left(\sigma^2 \right)^{b(n-1)}. \nonumber
\end{eqnarray}
The covariance matrix $\bm{\Sigma}$ is positive definite if the determinant is greater than zero. Therefore, the following restrictions are set
\begin{eqnarray}
\sigma^2  &>&0 \nonumber \\
\tau_b& > & -\sigma^2/n \nonumber \\
\tau_a &> & -\left(\tau_b/b + \sigma^2/(bn)\right). \nonumber
\end{eqnarray}
Note that these restrictions resemble the ones following from the expected sum of squares, SSA and SSB in Equation (17) and (19), respectively, which are restricted to be positive.

\section{C. BCSM with random interaction effects}
 The random treatment effect for the client, denoted as $\beta_{3ij}$, does not define a group effect, since each client $ij$ has only one post-intervention observation. Thus,  the  design matrix for the random treatment effect, $\mathbf{Z}_{i}$, is a diagonal matrix, with a one for each client in the EW condition at the post-intervention and a zero otherwise. The (conditional) MLM for counsellor $i$ can be presented as
 \begin{eqnarray}
\mathbf{y}_{i} & = & \beta_0 + \beta_1\mathbf{Treatment}_{i} + \beta_2\mathbf{Post}_{i} + \bm{\beta}_{3i}\mathbf{Z}_{i} + \bm{\beta}_{(i)}\left(\mathbf{I}_b\otimes\mathbf{1}_n\right) +
\alpha_i + \mathbf{e}_{i}, \nonumber \\
\alpha_i & \sim & N(0,\tau_a) \textrm{ (Counsellor)}   \nonumber \\
\bm{\beta}_{(i)} & \sim & N(0,\mathbf{I}_b\tau_b) \textrm{ (Clients)}  \nonumber \\
\bm{\beta}_{3i} & \sim & N(\beta_3,\mathbf{I}_{nb}\tau_c) \textrm{ (Interaction)}  \nonumber \\
\mathbf{e}_{i} & \sim & N(0,\mathbf{I}_{nb}\sigma^2). \nonumber
\end{eqnarray}
The covariance structure implied by the random effects for the clients of counsellor $i$ is given by
\begin{eqnarray}
\bm{\Sigma}_i & = & Var\left(\bm{\beta}_{3i}\mathbf{Z}_{i}\right) + Var\left(\alpha_i\mathbf{1}_{nb}\right) + Var\left(\bm{\beta}_{(i)}\left(\mathbf{I}_b\otimes\mathbf{1}_n\right)\right) + Var\left(\mathbf{e}_{i}\right) \nonumber \\
& = & \mathbf{Z}_{i}\mathbf{Z}^t_{i}\tau_c + \left(\mathbf{1}_{nb}\mathbf{1}^t_{nb}\right)\tau_a + \left(\mathbf{I}_b\otimes\mathbf{1}_n\right)\left(\mathbf{I}_b\otimes\mathbf{1}_n\right)^t\tau_b + \mathbf{I}_{nb}\sigma^2 \nonumber \\
& = & \mathbf{I}_{nb}\sigma^2  + \mathbf{Z}_{i}\mathbf{Z}^t_{i}\tau_c + \mathbf{J}_{nb}\tau_a + \left(\mathbf{I}_b\otimes\mathbf{J}_n\right)\tau_b \nonumber \\
& = & \mathbf{I}_{nb}\sigma^2 + \underbrace{diag(\mathbf{Z}_{i})\tau_c}_{\text{Interaction}} + \underbrace{\mathbf{J}_{nb}\tau_a}_{\text{Counsellor}} + \underbrace{\left(\mathbf{I}_b\otimes\mathbf{J}_n\right)\tau_b}_{\text{Client}}.
\end{eqnarray}
The posterior distribution of the residual variance $\sigma^2$ and random effect variance $\tau_c$ can be derived (Step 1). Consider the expected value of the sum of squares for the scores of clients in the AW condition,
\begin{eqnarray}
E\left(SS_{E_{AW}} \right) & = & E\left( \sum_{i=1}^a \sum_{j \in AW}\sum_{k=1}^n \left( y_{ijk} - \overline{y}_{ij.} \right)^2\right) \nonumber \\
& = & E\left( \sum_{i=1}^a \sum_{j \in AW}\sum_{k=1}^n \left( e_{ijk} - \overline{e}_{ij.} \right)^2\right) \nonumber \\
& = &  n_0\left(n-1\right)\sigma^2, \nonumber
\end{eqnarray}
where $n_0$ is the number of clients in the AW condition across all counsellors. Subsequently, variance parameter $\sigma^2$ has an inverse-gamma distribution with shape parameter $(g1 + n_0(n-1))/2$ and scale parameter $(g2 + SS_{E_{AW}})/2$. The posterior distribution of the variance parameter $\tau_c$ is based on the sum of squares of the post-intervention scores of the clients in the EW condition. The expected value is given by
\begin{eqnarray}
E\left(SS_{E_{EW}} \right) & = & E\left( \sum_{i=1}^a \sum_{j \in EW} \left( y_{ij2} - \overline{y}_{2} \right)^2\right) \nonumber \\
& = & E\left( \sum_{i=1}^a \sum_{j \in EW} \left( e_{ij2} - \overline{e}_{2} \right)^2\right) \nonumber \\
& = &  \left(n_1-1\right)\left(\sigma^2 + \tau_c\right), \nonumber
\end{eqnarray}
where $n_1$ is the number of post-intervention scores of clients in the EW condition, and $\overline{y}_2$ the average post-intervention score of all clients in the EW condition. The prior for $\sigma^2$ is an inverse-gamma with parameters $g_1$ and $g_2$. The prior for the $\tau_c$ is a shifted inverse gamma distribution, with the $\sigma^2$ as the shift parameter, and shape and scale parameter $g_1$ and $g_2$, respectively,
\begin{eqnarray}
p\left(\tau_c \mid \sigma^2 \right)  \propto \left(\tau_c + \sigma^2 \right)^{-g_1-1} \exp\left(\frac{-g_2}{\tau_c + \sigma^2} \right)\nonumber
\end{eqnarray}
and $\tau_c \ge -1/\sigma^2$. The posterior distribution of variance parameter $\tau_c$ is a shifted-inverse gamma distribution with shape parameter $(g1 + (n_1-1))/2$ and scale parameter $(g2 + SS_{E_{EW}})/2$.

The posterior distribution of $\tau_a$ and $\tau_b$ depend on the average residual variance (see Equation (17) and (19); step 2a and step 2b). With heteroscedastic error variances within a cluster $i$, a (pooled) average variance parameter is defined. The average residual variance can be defined using a pooled variance parameter. Consider the average residual variance,
\begin{eqnarray}
E\left(\overline{e}^2_{ij.}\right) & = & Var\left(\frac{e_{ij1} + e_{ij2}}{2}\right) = \left\{
  \begin{array}{ll}
  \sigma^2/2 & \textrm{ AW condition}\\
 \left(\sigma^2 +\tau_c/2\right)/2 & \textrm{ EW condition} \\
  \end{array}
\right. \nonumber
\end{eqnarray}
This expression is used to define the average residual variance in cluster $i$ using a pooled residual variance parameter. Let $n_{0i}$ and $n_{1i}$ define the number of clients in the AW and EW condition for counsellor $i$, respectively. It follows that,
\begin{eqnarray}
E\left(\overline{e}^2_{i..}\right) & = & \frac{n_{0i}\sigma^2/2 + n_{1i}\left(\sigma^2 +\tau_c/2\right)/2}{\left(n_{0i}+n_{1i}\right)^2} \nonumber \\
& = & \frac{\left(\frac{n_{0i}}{n_{0i}+n_{1i}}\right)\sigma^2 + \left(\frac{n_{1i}}{n_{0i}+n_{1i}}\right)\left(\sigma^2 +\tau_c/2\right)}{2\left(n_{0i}+n_{1i}\right)} \nonumber \\
& = & \frac{\tilde{\sigma}^2}{2\left(n_{0i}+n_{1i}\right)}  = \frac{\tilde{\sigma}^2}{nb}, \nonumber
\end{eqnarray}
with $n=2$, and $b=n_{0i}+n_{1i}$, and $\tilde{\sigma}^2$ the pooled residual variance parameter for cluster $i$. Finally, with $n_0$ and $n_1$ the total number of clients in the AW and EW condition, respectively, a general pooled residual variance parameter is defined using the weights $n_0/(n_0+n_1)$ and the $n_1/(n_0+n_1)$. This pooled variance parameter is used to define the shift parameter in the shifted-inverse gamma distribution of $\tau_a$ and $\tau_b$ (Step 3a and Step 3b).

\section{D. Extended Table 2: E-mail-counseling study.}

In Table \ref{extendedtable2}, the 95\% highest posterior density (HPD) intervals are added to Table 2 (in main paper) to provide more information about the posterior distribution of the BCSM parameters. In specific, the covariance parameters have skewed distributions, and more insight is provided by the HPDs about the possible range of plausible covariance values.

\begin{sidewaystable}[hbt!]
	\caption{Extended Table of the e-mail-counselling study: A BCSM analysis of the pre- and post-intervention data.}
	\begin{tabular}{lllclcrlrrlrr}
		\hline
		&                      &                      & LM ($\hat{M}$, $S.E.$)             &                      & \multicolumn{8}{c}{BCSM ($\hat{M}$, $SD$, 95\%HPD)}                                                                                         \\ \cline{4-4} \cline{6-13}
		\multicolumn{1}{c}{} & \multicolumn{1}{c}{} & \multicolumn{1}{c}{} & M0  & & \multicolumn{2}{c}{M1}    & \multicolumn{1}{c}{} & \multicolumn{2}{c}{M2} & \multicolumn{1}{c}{} & \multicolumn{2}{c}{M3} \\ \cline{4-4} \cline{6-7} \cline{9-10} \cline{12-13}
		\multicolumn{2}{l}{\textit{Fixed effect}}   &                      &                                    &                      &                                    &                      & \multicolumn{1}{c}{}   &                      & \multicolumn{1}{c}{}   & & \\ \cline{1-2}
		Intercept            & $\beta_{0}$          &                      & \multicolumn{1}{r}{21.78 (0.91)}   &                      & \multicolumn{1}{r}{21.72 (0.84)}   & \multicolumn{1}{r}{(20.06,23.34)}   &                     & 21.67 (0.79)          & \multicolumn{1}{r}{(20.11,23.21)}  &                      & 21.68 (0.80)    & \multicolumn{1}{r}{(20.14,23.29)}       \\
		Treatment            & $\beta_{1}$          &                      & \multicolumn{1}{r}{$-$0.29 (1.29)} &                      & \multicolumn{1}{r}{$-$0.16 (1.29)} & \multicolumn{1}{r}{($-$2.65,2.39)}  &
		& $-$0.08 (1.29)    & \multicolumn{1}{r}{($-$2.61,2.44)}     &                      & $-$0.09 (1.31)     & \multicolumn{1}{r}{($-$2.65,2.47)}    \\
		Post                 & $\beta_{2}$          &                      & \multicolumn{1}{r}{$-$4.04 (1.28)} &                      & \multicolumn{1}{r}{$-$4.06 (1.29)} & \multicolumn{1}{r}{($-$6.59,$-$1.56)} &
		& $-$4.03 (0.99)      & \multicolumn{1}{r}{($-$6.03,$-$2.13)}   &                      & $-$4.05 (1.01)    & \multicolumn{1}{r}{($-$6.08,$-$2.11)}     \\
		Interaction          & $\beta_{3}$          &                      & \multicolumn{1}{r}{$-$1.36 (1.81)} &                      & \multicolumn{1}{r}{$-$1.33 (1.82)} & \multicolumn{1}{r}{($-$4.97,2.18)} &                      & $-$1.37 (1.42)    & \multicolumn{1}{r}{($-$4.10,1.44)}     &                      & $-$1.35 (1.43)   & \multicolumn{1}{r}{($-$4.17,1.48)}      \\
		&                      &                      &                                    &                      &                                    &                      & \multicolumn{1}{c}{}   &                      & \multicolumn{1}{c}{}   \\
		\multicolumn{2}{l}{\textit{Random effects}} &                      &                                    &                      &                                    &                      & \multicolumn{1}{c}{}   &                      & & & \multicolumn{1}{c}{}   \\ \cline{1-2}
		Residual             & $\sigma^{2}$         &                      & 37.04                              &                      & \multicolumn{1}{r}{37.79 (4.03)}  & \multicolumn{1}{r}{(30.03,45.71)} &                      & 21.73 (3.32)  &  \multicolumn{1}{r}{(15.70,28.44)} &                 & 22.08 (4.89)  &  \multicolumn{1}{r}{(13.88,32.03)}         \\
		Counsellor           & $\tau_{a}$           &                      &                                    &                      & \multicolumn{1}{r}{$-$0.68 (0.44)} & \multicolumn{1}{r}{($-$1.15,0.06)} &                      & $-$1.12 (0.49)  & \multicolumn{1}{r}{($-$1.86,$-$0.27)}         &                      & $-$1.07 (0.51)   & \multicolumn{1}{r}{($-$1.84,$-$0.27)}        \\
		Client               & $\tau_{b}$           &                      &                                    &                      &                                    &        &              & 15.83 (4.52)   &  \multicolumn{1}{r}{(7.43,24.98)}        &                      & 15.54 (4.74)  &   \multicolumn{1}{r}{(6.73,25.13)}        \\
		Interaction          & $\tau_{c}$           &                      &                                    &                      &                                    &         &             & \multicolumn{1}{c}{}   &          &            & 6.28 (7.79)     & \multicolumn{1}{r}{($-$9.45,21.71)}         \\ \hline
	\end{tabular}
	\label{extendedtable2}
\end{sidewaystable}

\printbibliography

\end{document}